\documentclass[sn-mathphys,Numbered]{sn-jnl}

\usepackage{graphicx}
\usepackage{float}
\usepackage{multirow}%
\usepackage{amsmath,amssymb,amsfonts}
\usepackage{amsthm}
\usepackage{mathrsfs}
\usepackage[title]{appendix}
\usepackage{xcolor}
\usepackage{textcomp}
\usepackage{manyfoot}%
\usepackage{booktabs}%
\usepackage{algorithm}
\usepackage{mathtools}
\usepackage{algorithmicx}%
\usepackage{algpseudocode}%
\usepackage{listings}%
\usepackage{lmodern}
\usepackage[capitalize]{cleveref}
\usepackage{anyfontsize}
\usepackage{hyperref}
\usepackage{cleveref}
\usepackage{mathtools}

\begin{document}

\title{The Spectroscopy of the 2+1 Dimensional Analog Black Hole in Photon-Fluid Model}

\author[1]{David Senjaya} \email{davidsenjaya@protonmail.com} 
\affil[1]{Department of Physics, Faculty of Science, Mahidol University, Bangkok 10400, Thailand}

\author[2]{Supakchai Ponglertsakul}
\email{supakchai.p@gmail.com}
\affil[2]{Strong Gravity Group, Department of Physics, Faculty of Science, Silpakorn University, Nakhon Pathom 73000, Thailand}


\maketitle
\begin{abstract}

In this paper, we explore quasibound states (QBS), scalar cloud, Hawking radiation, superradiance, and greybody factor of relativistic massive phonon modes in a photon-fluid rotation black hole. We investigate quasibound states and scalar clouds using exact eigensolutions to the analog Klein-Gordon equation in the analog black hole background and revisit the Wentzel-Kramers-Brillouin (WKB) upper bound on the scalar clouds' energy ratio. Using the obtained exact radial solution, we use the Damour-Ruffini method to calculate the power spectrum of the analog black hole's Hawking radiation. We then use the analytical asymptotic matching technique (AAM) to investigate the analog black hole's superradiance for low energy massive photon  scattering, resulting in the analytical amplification factor and the greybody factor formulas of the analog black hole. We discover that the analog black hole in the photon-fluid model is superradiant with an energy range of $\varpi < \omega<m_\ell\Omega_H$. As a result, the greybody factors are negative for co-rotating modes in the superradiant regime.

\end{abstract}

\section{Introduction}
Analytical investigation into the solutions to a relativistic wave equation in black hole spacetime is critically important to study the black hole's response to a certain perturbation. There are various applications in black hole physics that benefit from having an exact solution to the wave equation. For instance, several works \cite{Lei,Siqueira,Noda,senjaya1,senjaya2,Senjaya:2024rse,Senjaya:2025bbp} using an exact solution to the wave equation to investigate the greybody factor, scalar cloud, quasinormal modes, entropy and area quantization, quasibound states, and superradiance in various black hole spacetimes. In general, bounded scalar perturbations are represented by a discrete spectrum of complex frequencies known as the quasiresonance frequency, with the real component determining the oscillation timescale and the imaginary part determining the exponential decay (stability) or growth (instability). As a result, quasiresonance frequency is vital in understanding black hole stability under certain perturbations.

Moreover, due to the lack of experimental access to the Planck scale, researchers are looking for indirect ways to understand quantum gravity. Quantization of scalar field in black hole spacetimes are crucial, similar to how atomic models helped establish quantum theories \cite{Keshet}. In addition, there has been a paucity of experimental feedback in studies of crucial phenomena in general relativity and quantum field theory in curved spacetime. Alternative attempts have been made to identify non-relativistic systems that can be empirically tested in the laboratory that is known as the analogue black holes, such as Bose-Einstein condensates, electromagnetic wave guides, graphene, optical black hole, acoustic black hole, and ion rings \cite{Mathie,Visser,Unruh1, Kandemir,Mann}. Analogue gravity models are gaining popularity because they may be used to test many elements of relativistic quantum field theory in a curved spacetime. 

Here, we will particularly focus on the photon-fluid model, a non-linear optical system described by hydrodynamic equations of an interacting Bosonic gas \cite{roton}. In \cite{PhysRevA.100}, a photon-fluid with both local and non-local interactions is explored, revealing that phonons acquire a limited mass and the governing equation is similar to the massive Klein-Gordon equation in a (2+1)-dimensional curved space-time. With appropriate vortex fluxes, this system may generate both quasibound and stationary phonon states \cite{Ciszak}. These are the acoustic version of quasibound states and scalar clouds near Kerr black holes \cite{vier21,Senjaya3,Senjaya:2024blm,Senjaya:2024gpb,Senjaya:2024uqg}. In \cite{Ciszak}, numerical quasibound states and scalar clouds are reported for the first time, and the WKB results of the photon fluid black hole's scalar clouds are found in \cite{PhysRevD.103,Hod:2021pkw}. Two years later, Senjaya et al. \cite{Senjaya:2024xqm} discovered the exact quasibound states of the photon fluid black hole following the development on the Heun functions research.

Therefore, we consider 2+1 dimensional rotating analog black hole. In Sec \ref{sect:BH}, we provide a brief review on the photon-fluid model and discuss some basic properties of the analog black hole. The analog 2+1 dimensional Klein-Gordon equation is derived in Sec \ref{sect:KG}. In Sec \ref{sect:qbs}, the exact quasibound sates are obtained and is implemented to investigate the black hole analog's scalar clouds configuration. Also, the upper bound of the scalar clouds mass to angular momentum ratio is calculated and compared to the previously known WKB results \cite{PhysRevD.103,Hod:2021pkw}. Additionally in Sec \ref{sect:BE}, the analog Hawking's radiation is explored via 
the Damour-Ruffini method \cite{Damour}. We use the exact radial solution of the Klein-Gordon equation to derive the boson distribution function near the black hole's event horizon and calculate the radiation power of the Hawking radiation.

The quest for black hole superradiance amplification involves calculating the transmission and reflection amplitudes of linear perturbations from infinity. In this inquiry, we utilize the analytical asymtotic matching (AAM) approach to derive the amplification factor in Sec \ref{sect:suprad}. 
To close this work, We use the obtained transmission and reflection amplitudes expressions to further calculate the black hole analog's greybody factor in Sec \ref{sect:gbf}.  It is worth noting that the superradiance scattering and greybody factor of the photon-fluid black hole are calculated analytically for the first time in this work. Lastly, we summarize our findings in Sec \ref{sect:conclud}.

\section{The Photon-Fluid Model}\label{sect:BH}
A photon-fluid system is a non-linear optical system described by hydrodynamic equations of interacting Bosonic gas \cite{roton}. The photon-fluid system, like the exciton-polariton system and Bose-Einstein condensation of photons in an optical microcavity \cite{Klaers}, belongs to the family of the so-called quantum fluids of light \cite{Carusotto}. The investigation of a photon-fluid with both local and non-local interactions in \cite{PhysRevA.100} demonstrates that the photons acquire mass and propagate in a similar way with a massive scalar field, whose dynamics is described by a massive Klein-Gordon equation, in a (2+1)-dimensional curved space-time. Both stationary and quasi stationary bound states of phonons can exist in this system \cite{Ciszak} given appropriated vortex fluxes are present. Therefore, the photon-fluid-boson system can be regarded as an acoustic counterpart of scalar clouds and quasibound states that bound to a rotating black hole. 

The current work focuses on the deeper investigation to the quasibound states, scalar clouds, statistic near the analog black hole horizon, greybody factor, and superradiance, whereas in our previous work \cite{Senjaya:2024xqm}, the exact quasibound states formula are derived and comparison to the numerical results have been presented. In \cite{PhysRevD.103,Hod:2021pkw}, scalar clouds in photon fluids have been derived in  eikonal large frequency regime requiring exact method for other frequency regime.

Any inhomogeneous neutrally stable nonlinear system, possessing linear and non-linear refractive indexes, $n_0, n_2$, respectively, may generally propagate its linear elementary excitations over an effective curved space-time with local speed $v^2(r)=\frac{n_2 I_0}{n_0^3\mu_0\epsilon_0}$. We consider the first order fluctuations of to the background field as follows,
\begin{equation}
    E=\sqrt{I(r)} e^{i\left(m\theta-2\pi \sqrt{\frac{r}{r_0}}\right)},
\end{equation}
where,
\begin{equation}
I(r)=I_0+\varepsilon I_1(r)+... \ \ \ , \ \ \ |\varepsilon|<<1,
\end{equation}
and with $I_0$ is the electromagnetic field  background intensity of the helical wave, $E$ \cite{31}, $m$ is angular integer, $\epsilon_0, \mu_0$ are vacuum electric permittivity and magnetic susceptibility, and $r$ is radial coordinate of the system. When a beam with wavelength $\lambda$, possessing radial and angular velocity consecutively, 
\begin{equation}
   v_r(r)=-\frac{\lambda}{2n_0\sqrt{\mu_0\epsilon_0r_0r}} \ \ \ , \ \ \ v_\theta(r)=\frac{\lambda m}{2\pi n_0 r\sqrt{\mu_0\epsilon_0}},
\end{equation}
is injected, analog ergoregions and event horizon are formed. The parameter $r_0$ is experimental parameter controlling the radial phase dependence of the beam \cite{Vocke:2018xwb}.  

Furthermore, a Klein-Gordon-like equation can be derived by linearization around the background state, governs the dynamics of phonons propagating in an inhomogeneous photon flow analogous to relativistic scalar fields in curved space-time. The equivalent space-time metric is then extracted by comparing the governing equation with the Klein-Gordon equation in 2+1 dimension resulting in this following expression \cite{PhysRevA.100}, 
\begin{gather}
ds^2=-f(r)d(ct)^2+\frac{dr^2}{g(r)}-2\omega_H r_H^2 d\theta d(ct) +r^2 d\theta^2, \label{metric}\\ 
g(r)=1-\frac{r_H}{r}=\Delta,\\
f(r)=1-\frac{r_H}{r}-\frac{\omega^2_H r_H^4}{r^2}=\Delta-\frac{\omega^2_H r_H^4}{r^2},
\end{gather}
where the dimensionless angular velocity at the horizon, $\omega_H$, is given by,
\begin{equation}
    \omega_H\equiv \frac{v_\theta(r_H)}{r_H v(r_H)n_0}=\frac{m \xi}{\pi r_H^2}. \label{angmom}
\end{equation} 
The radius of the analog black hole's horizon, denoted by $r_H=\frac{\xi^2}{r_0}$, with $\xi=\frac{\lambda}{2\sqrt{n_0n_2I_0}}$ \cite{31}. The horizon is determined from  $g(r_H)=0$ and can be geometrically understood as a circular ring where the fluid's inward radial flow velocity equals to the characteristic speed, $v_r(r_H)=-v(r_H)$ \cite{Ciszak}. There is a region where $f(r)<0$, called ergoregion with radius,
\begin{equation}
    r_{ergo}=\frac{r_H}{2}+\frac{r_H}{2}\sqrt{1+4\omega_H^2r_H^2}.
\end{equation} 

Note that the presence of the cross term $d(ct)d\phi$ represents the existence of frame dragging effect that disappears at the limit $\omega_H \to 0$.

\section{The Analog Klein-Gordon Equation}\label{sect:KG}

In the photon-fluid model's sound wave regime, it has been reported in \cite{Ciszak} that the dynamics of acoustic density fluctuations in the presence of vortex flows is described by an analog massive Klein-Gordon equation in a non-trivial (2 + 1)-dimensional curved spacetime presented as the following, 
\begin{gather}
\left[-\frac{1}{\sqrt{-g}}{\partial }_\mu\left(\sqrt{-g}g^{\mu \nu}{\partial }_\nu\right)+\varpi^2\right]\psi=0,\label{KGEq}
\end{gather}
where $\psi$ represents first-order fluctuations of the electromagnetic field. The analog rest energy, $\varpi$, has generic relation to the medium refractive index, as follows {\cite{Ciszak}, 
\begin{equation}
  {\varpi}= \sqrt{\frac{\alpha\beta I_0}{n_0^3\epsilon_0\mu_0}}, 
\end{equation}
where $\alpha, \beta$ comes from the optical response of the non-linear medium, both local Kerr, $n_2$, and nonlocal thermo-optical nonlinearities, $n_{Th}(T)$, as a function of temperature \cite{Ciszak},
\begin{gather}
   \Delta n=n_2|E|^2+\int \frac{\alpha \beta |E|^2}{4\pi |\vec{r}-\vec{r}'|}dVol'=(n_2+n_{Th})|E|^2,\\
   -\nabla^2 n_{Th}=\alpha \beta |E|^2 \ \ \ , \ \ \ \beta=\frac{\partial n_{Th}}{\partial T}>0.
\end{gather}
Also note that classically, the following dispersion relation holds \cite{Ciszak},
\begin{equation}
    \omega^2=\varpi^2+v^2 \left(\frac{2\pi}{\lambda}\right)^2 +v^2 r_H r_0 \left(\frac{4\pi^2}{\lambda^4}\right).\label{dispersion}
\end{equation}
The Klein-Gordon like equation above is true in $\frac{1}{\lambda^4}<<1$ regime, where the dispersion relation \eqref{dispersion} reduces to a regular relativistic dispersion relation for massive bosons.

Substituting the photon-fluid metric \eqref{metric} into the analog massive Klein-Gordon equation \eqref{KGEq}, followed by applying separation ansatz,
\begin{equation}
\psi(ct,r,\theta)=e^{-i\frac{E}{\hbar c}ct}e^{im_\ell\theta}\frac{R(r)}{\sqrt{r}}. \label{ansatz}
\end{equation}
The radial equation in the terms of new dimensionless radial variable $x=\frac{r}{r_H}$ is obtained  (See Appendix \ref{Appendix0} for the derivation)
\begin{equation}
\Delta {\partial }_x\left(\Delta {\partial }_x\right)R\left(x\right)+\left[\Delta \left(\frac{\Delta }{4x^2}-\frac{1}{2x^3}-\frac{m_\ell^2}{x^2}-\varpi^2\right)+{\left(\omega -\frac{m_\ell{\Omega }_H }{x^2}\right)}^2\right]R\left(x\right)=0, \label{mainradeq}    
\end{equation}
where $m_\ell$ is the harmonic quantum number and we have defined $\omega=\frac{Er_H}{\hbar c}$. The following equation is obtained by multiplying the whole radial equation \eqref{mainradeq} by ${\Delta }^{{-}{2}}$ and expanding the double derivatives in the first term,
\begin{gather}
{\partial }^{2}_xR+p\left(x\right){\partial }_xR+q\left(x\right)R=0, \label{radialeq}
\end{gather}
with,
\begin{align}
p\left(x\right)&=\frac{1}{x-1}-\frac{1}{x},\\
q\left(x\right)&=\left(\frac{x}{x-1}\right)\left(\frac{x{-}{1}}{4x^3}-\frac{1}{2x^3}-\frac{m_\ell^2}{x^2}-\varpi^2 \right)+{\left(\frac{x}{x-1}\right)}^2{\left(\omega -\frac{m_\ell{\Omega }_H}{x^2}\right)}^2.
\end{align}
One can utilize these following fractional decomposition formulas,
\begin{align}
\frac{1}{x\left(x-1\right)}&=\frac{1}{x-1}-\frac{1}{x},\\
\frac{{1}}{x^2\left(x-1\right)}&=\frac{1}{x-1}-\frac{1}{x}-\frac{1}{x^2},\\
\frac{{1}}{x^2{\left(x-1\right)}^2}&=-\frac{2}{x-1}+\frac{1}{{\left(x-1\right)}^2}+\frac{2}{x}+\frac{1}{x^2},\\
\frac{x}{x-1}&=\frac{1}{x-1}+1,    
\end{align}
to express $q(x)$ as follows,
\begin{multline}
q\left(x\right)=\left({\omega }^{2}-\varpi^2 \right)+\frac{1}{x}\left(\frac{1}{2}+m_\ell^2\left(1+2{\Omega }^2_H\right)\right)\\
+\frac{1}{x-1}\left(-\frac{1}{2}-m_\ell^2\left(1+2{\Omega }^2_H\right)-\varpi^2 +2{\omega }^{2}\right)\\+\frac{1}{x^2}\left(\frac{3}{4}+m_\ell^2{\Omega }^2_H\right)+\frac{1}{{\left(x-1\right)}^2}{\left(\omega -m_\ell{\Omega }_H\right)}^2. 
\end{multline}
For notational convenience, we define a new radial variable $z=-\left(x-1\right)$, which shifts the domain of interest to $-\infty < z\leq 0$}. This is because we are interested in finding solutions in the region outside the horizon, i.e. $r_H\leq r<\infty$ or equivalently $1\leq x<\infty $. The radial equation in terms of $z$ is,
\begin{align}
{\partial }^{2}_zR+p\left(z\right){\partial }_zR+q\left(z\right)R=0,\label{radialz}
\end{align}
with
\begin{align}
p\left(z\right)&=\frac{1}{z}-\frac{1}{z-1}, \label{pz}\\
q\left(z\right)&=-\left(\varpi^2-\omega ^2\right)-\frac{1}{z-1}\left(\frac{1}{2}+m_\ell^2\left(1+2{\Omega }^2_H\right)\right) \nonumber \\
&~~~~~-\frac{1}{z}\left(-\frac{1}{2}-m_\ell^2\left(1+2{\Omega }^2_H\right)-\varpi^2 +2{\omega }^{2}\right) \nonumber \\
&~~~~~+\frac{1}{{\left(z-1\right)}^2}\left(\frac{3}{4}+m_\ell^2{\Omega }^2_H\right)+\frac{1}{z^2}{\left(\omega -m_\ell{\Omega }_H\right)}^2. \label{qz}      
\end{align}
Following Appendix \ref{AppendixB}, we obtain the normal form of the above radial equation as follows, 
\begin{gather}
{\partial }^{2}_zY\left(z\right)+K\left(z\right)Y\left(z\right)=0,    \label{radialnormalform1}
\end{gather}
where $Y\left(z\right)=z^{1/2}\left(1-z\right)^{-1/2}R(z)$ and  
\begin{align}
K\left(z\right)&=-\frac{1}{2}{\partial }_zp\left(z\right)-\frac{1}{4}p^2\left(z\right)+q\left(z\right), \nonumber \\
&=-\left(\varpi^2-{\omega }^{2} \right)-\frac{1}{z-1}\left(m_\ell^2\left(1+2{\Omega }^2_H\right)\right)-\frac{1}{z}\left(2{\omega }^{2}-m_\ell^2\left(1+2{\Omega }^2_H\right)-\varpi^2 \right) \nonumber \\
&~~~~~+\frac{1}{{\left(z-1\right)}^2}\left(m_\ell^2{\Omega }^2_H\right)+\frac{1}{z^2}\left(\frac{{1}}{{4}}{+}{\left(\omega -m_\ell{\Omega }_H\right)}^2\right). \label{radialnormalform2}
\end{align}
We determine the exact radial solution of the radial Klein-Gordon equation in the vortex flow of photons by comparing the normal form of the radial equation above with the normal form of the Confluent Heun's differential equation in Appendix \ref{AppendixC} and obtain the following results,
\begin{gather}
    R=e^{\frac{1}{2}\alpha_\pm z}z{\left(z-1\right)}^{\frac{1}{2}\gamma_\pm}\times \left[R_{\pm}^N z^{\frac{\beta_\pm}{2}}\operatorname{HeunC}(\alpha_\pm,\beta_\pm,\gamma_\pm,\delta,\epsilon,z)\right], \label{radialsol}
\end{gather}
where $R_{\pm}^N$ is the radial wave function's normalization constant. Here, we use compact notation defined in \eqref{HeunSol}. The parameters of the Confluent Heun function are solved algebraically as follows,
\begin{align}
\alpha_\pm &=\pm 2\sqrt{\varpi^2-{\omega }^{2}},\\
\beta_\pm &=\pm2i\left(\omega -m_\ell{\Omega }_H \right), \label{beta}\\
\gamma_\pm &=\pm i\sqrt{4m_\ell^2{\Omega }^2_H-1}=\pm i\left|\gamma\right|,\\
\delta &=\varpi^2 -2{\omega }^{2},\\
\eta &=\frac{1}{2}-m_\ell^2\left(1+2{\Omega }^2_H\right)-\varpi^2 +2{\omega }^{2}.
\end{align}

\section{Quasibound States and Scalar Clouds}\label{sect:qbs}
Now, we are going to discuss the analog quasibound states and the scalar clouds in the photon-fluid model. The quasibound states are quantized relativistic bound states that exist in the black hole's gravitational potential well outside its event horizon. In contrast to real bound states, quasibound states seep into the black hole, resulting in complex-valued frequencies in the spectrum. The system's stability is determined by the imaginary component of the spectrum \cite{Vieira:2021doo, Senjaya:2024blm}. Mathematically, the quasibound states have these following nature,
\begin{itemize}
    \item The real part of the quasibound state relativistic energy, $\operatorname{Re}(\omega)$, is always less than its rest energy $\varpi$,
    \begin{equation}
        \varpi>\operatorname{Re}(\omega).
    \end{equation}
    \item Very near to the black hole's event horizon $r=r_H$, the radial wave function behaves as a purely ingoing wave. The near horizon limit of \eqref{radialsol} is obtained as follows,
    \begin{equation}
    R(r\to r_H) \sim \left(r-r_H\right)^{\frac{\beta_\pm}{2}},
    \end{equation}
    where only $\beta_-$ mode fulfills the quasibound states' boundary condition.   
     
     \item Far away from the black hole's horizon, $r \to \infty$, the radial wave function is exponentially decaying. It is straightforward to get the far field limit of \eqref{radialsol} as follows,
     \begin{align}
    R(r\to \infty)&\sim e^{\frac{1}{2}\alpha_\pm z}{z}^{\frac{\gamma_\pm+\beta_\pm}{2}+1} \operatorname{HeunC}(\alpha_\pm,\beta_\pm,\gamma_\pm,\delta,\epsilon,z)\\
    &\sim e^{-\frac{1}{2}\alpha_\pm \frac{r}{r_H}}{r}^{\frac{\gamma_\pm+\beta_\pm}{2}+1} \operatorname{HeunC}\left(\alpha_\pm,\beta_\pm,\gamma_\pm,\delta,\epsilon,-\frac{r}{r_H}\right),
\end{align}
   where $r_H \leq r < \infty$. Note that after quantization, the $\operatorname{HeunC}$ function becomes a polynomial. Therefore, only $\alpha_+$ satisfies the quasibound states' boundary conditions. 
\end{itemize}

Only two modes-those with $\alpha_+,\beta_-,\gamma_\pm$-satisfy the boundary requirements out of the eight modes $\alpha_\pm,\beta_\pm,\gamma_\pm$. By further choosing ingoing wave condition at $r=0$, only the $\alpha_+,\beta_-,\gamma_-$ mode remains \cite{vier21}.

The termination condition of the Frobenius series expansion of the radial wave function yields the radial quantization condition. By using the confluent Heun polynomial condition \eqref{HeunPol}, we get the exact quantized energy formula of the quasibound states as follows,
\begin{gather}
\frac{\varpi^2 -2{\omega }^{2}}{2\sqrt{\varpi^2 -{\omega }^{2}}}-i\left(\omega -m_\ell{\Omega }_H+\frac{1}{2}\sqrt{4m_\ell^2{\Omega }^2_H-1}\right)=-n \ \ , \ \ n=1,2,3,... \ \label{exactenergy}.
\end{gather}
Furthermore, redefining the radial quantum number by $N=n+\ell+1$ \cite{Yang,Yang22} is another way to describe the radial quantum number, $n$, in terms of the total quantum number, $N$.

\begin{figure}[h]
    \centering
    \includegraphics[scale=0.6]{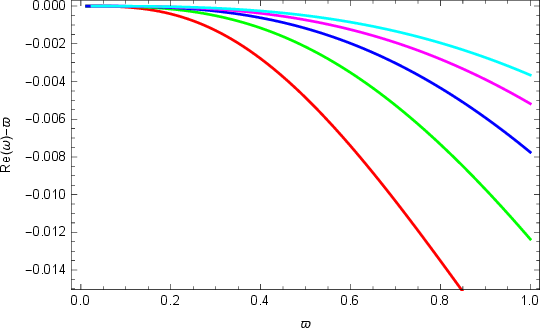}
    \includegraphics[scale=0.6]{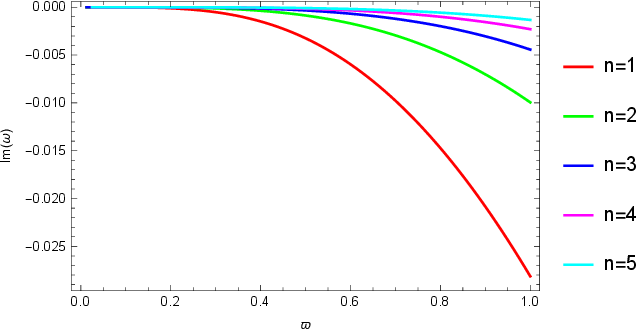}
    \caption{Profile of quasibound states frequencies with respect to the scalar mass $\varpi$ with fixed $\Omega_H=0.7,m_{\ell}=0$.} \label{QBS}
\end{figure}

In Fig. \ref{QBS}, we present ground state and four first excited states of the massive scalar's quasibound states profile with respect to the scalar mass $\varpi$ with $m_\ell=0$. Note that unlike Kerr black holes, the analog black hole's spin in the photon-fluid model always appears coupled to $m_\ell$. Therefore, the quasibound state frequency is insensitive to $\Omega_H$ for circular state $m_\ell=0$. We observe that as $n$ increases, both of quantity $\operatorname{Re}(\omega)-\varpi$ and $\operatorname{Im}(\omega)$ decrease (in magnitude for $\operatorname{Im}(\omega)$). 

\begin{figure}[h]
    \centering
    \includegraphics[scale=0.35]{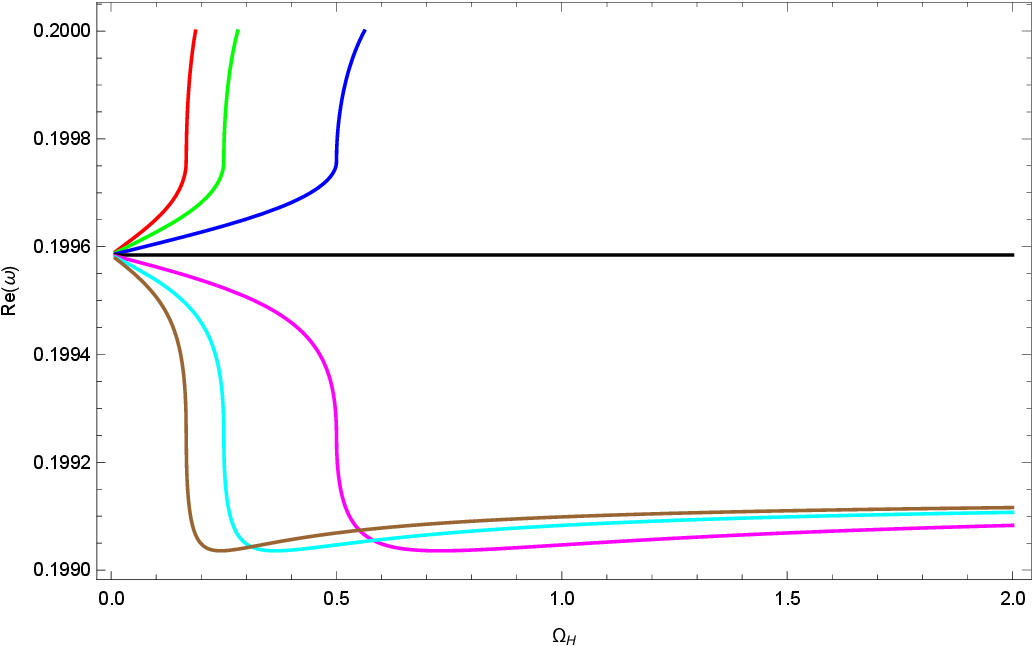}
    \includegraphics[scale=0.35]{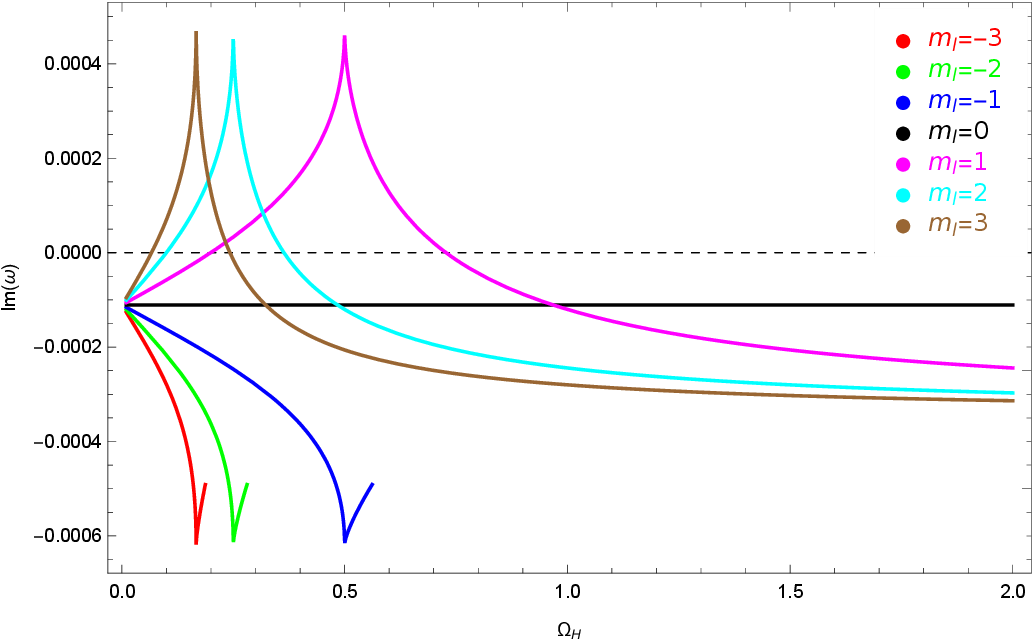}\\
    \includegraphics[scale=0.35]{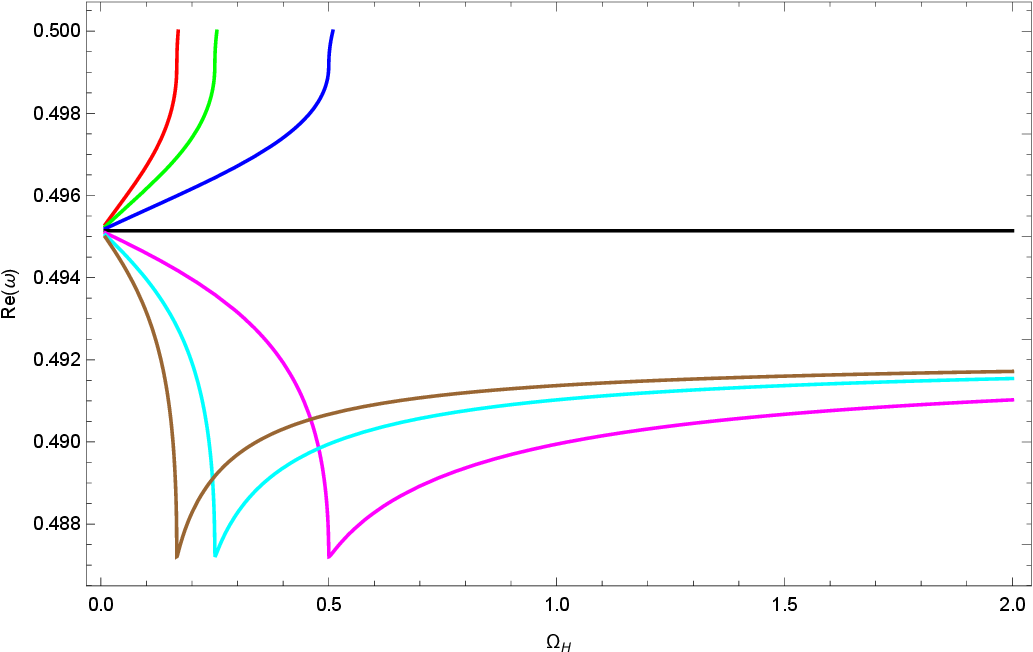}
    \includegraphics[scale=0.35]{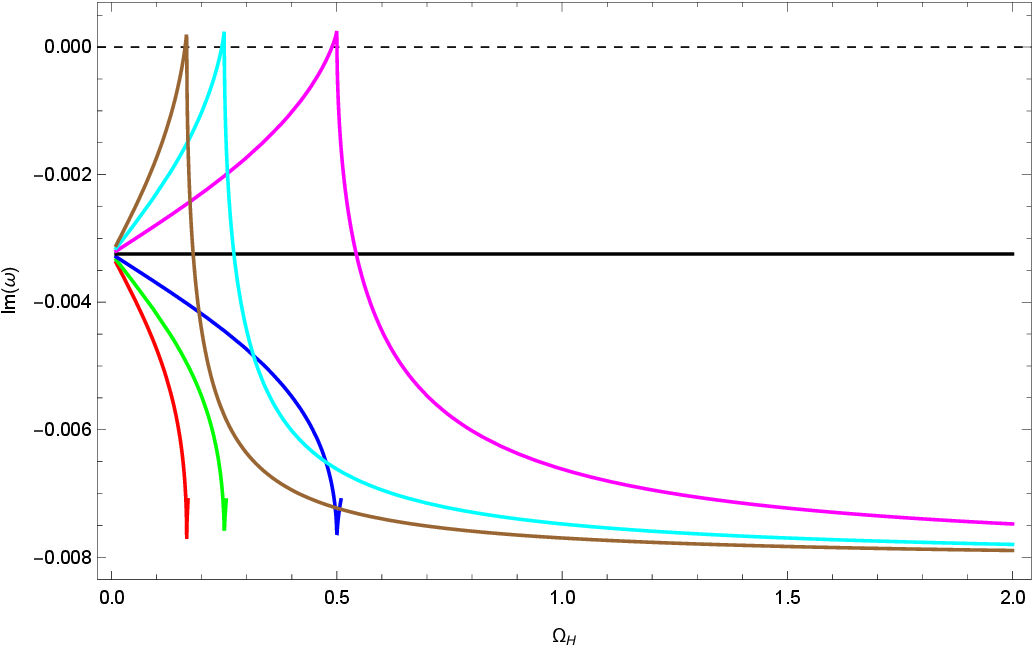}\\
    \includegraphics[scale=0.35]{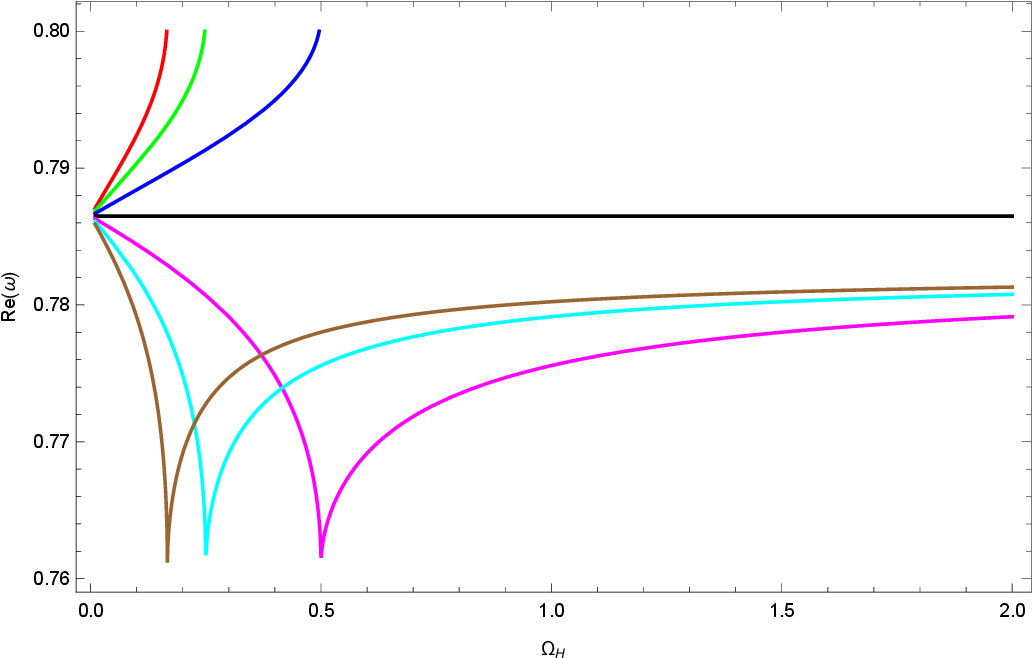}
    \includegraphics[scale=0.35]{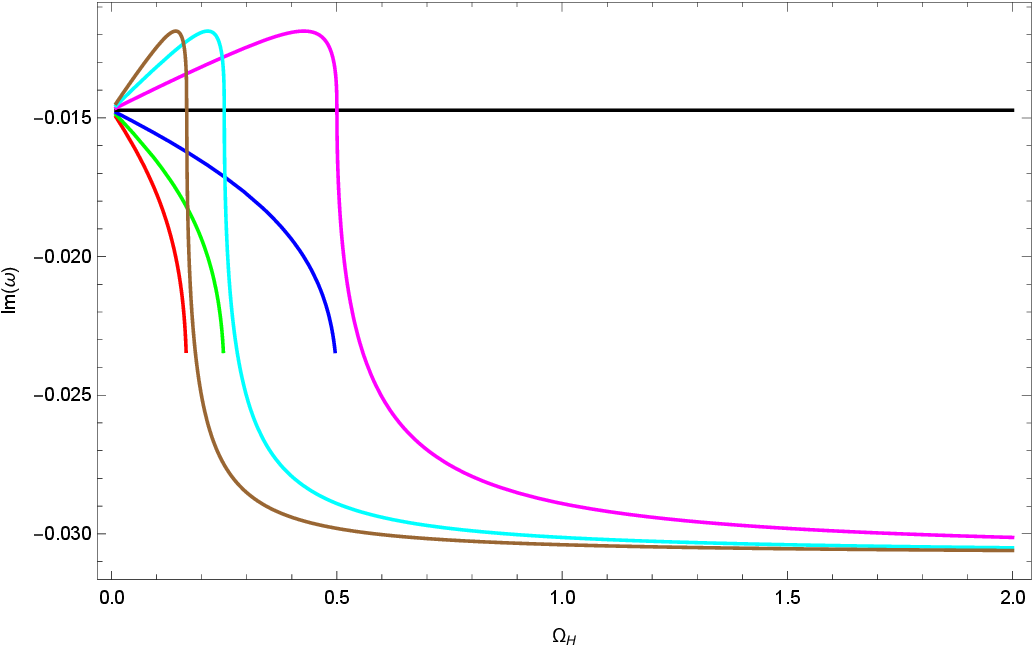}
    \caption{Profile of quasibound states frequencies with respect to the analog black hole spin $\Omega_H$ with fixed $\varpi=0.2,0.5,0.8$ (from top row to bottom row) for various value of $m_\ell$.} \label{QBS2}
\end{figure}

We  explore the effect of black hole spin $\Omega_H$ on the quasibound state frequencies with $n=1$ in Fig.~\ref{QBS2}. The rotation feature of analog black hole disrupts azimuthal degeneracy, i.e., bound states with different azimuthal numbers $m_\ell$ have different eigenfrequencies. Consequently, there is a significant difference in the behavior of the co-rotating ($m_\ell>0$) and counter-rotating states ($m_\ell<0$). The real (imaginary) part is shown as a function of black hole spin as displayed in the left (right) column of this figure. The real part exhibits generic feature such that for the counter-rotating states, the real frequency increases with black hole spin. At each fixed negative value of $m_\ell$, there is maximum value of $\Omega_H$ where quasibound state exists i.e., $\varpi-\operatorname{Re}(\omega)>0$. For instance, when $m_\ell=-3$, the maximum $\Omega_H$ is at $0.18$. Beyond this value, quasibound state no longer exists. As $m_\ell$ increases (less negative), the maximum value of $\Omega_H$ is also increases. For positive $m_\ell$, the real part drops at small $\Omega_H$ and later inclines before approaching a certain value at large $\Omega_H$.

Moreover, the imaginary part of $\omega$ also displays interesting behavior with black hole spin. First of all, counter-rotating quasibound state frequency curves render negative imaginary part. Conversely, the imaginary part of the co-rotating quasibound state frequency increases with $\Omega_H$ until a certain value and then begins to decline. More interestingly, at small $\varpi$, i.e. ($\varpi=0.2$) (top row), the imaginary part of $\omega$ can be positive before decreasing to negative value at large $\Omega_H$. Note that the existence of states with $\operatorname{Im}(\omega)>0$ depends on $\varpi$, i.e. the larger $\varpi$, the lower the chance of getting $\operatorname{Im}(\omega)>0$ for the co-rotating states. For large $\varpi$, i.e. ($\varpi=0.8$) (last row), there are no states with $\operatorname{Im}(\omega)>0$. Lastly, we also show that the quasibound state frequencies are independent of $\Omega_H$ when $m_\ell=0$ as illustrated by the horizontal lines. The existence of quasibound states with positive $\operatorname{Im}(\omega)$ indicates that scalar fields around the photon-fluid system experience superradiant instabilities. Superradiant boundstates, however, are unable to radiate and escape to infinity. These confined exponentially amplified modes are well-known as the black hole bomb \cite{Liu:2024qso}. 


\begin{figure}[h]
    \centering
    \includegraphics[scale=0.6]{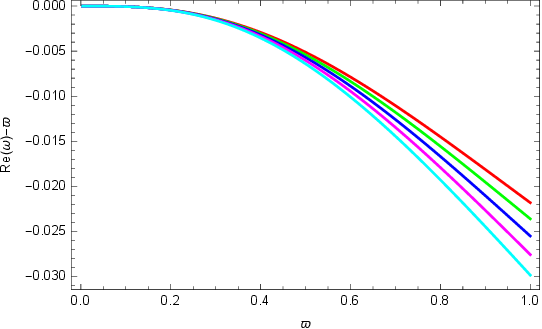}
    \includegraphics[scale=0.6]{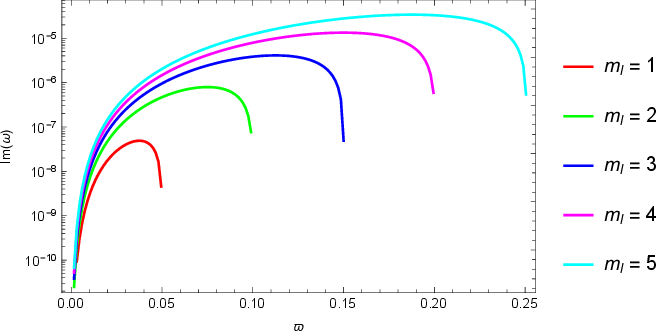}\\
     \includegraphics[scale=0.6]{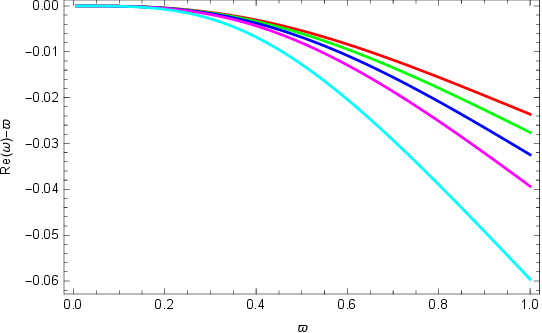}
    \includegraphics[scale=0.6]{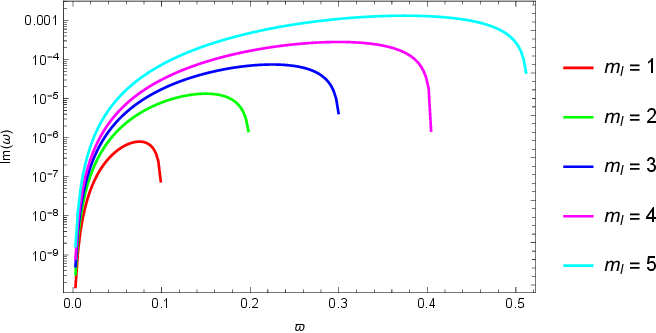}\\
     \includegraphics[scale=0.6]{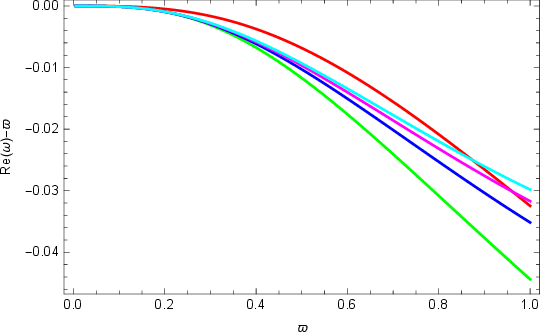}
    \includegraphics[scale=0.6]{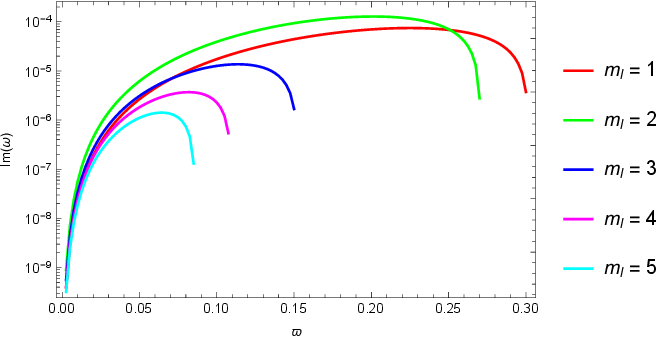}\\
     \includegraphics[scale=0.6]{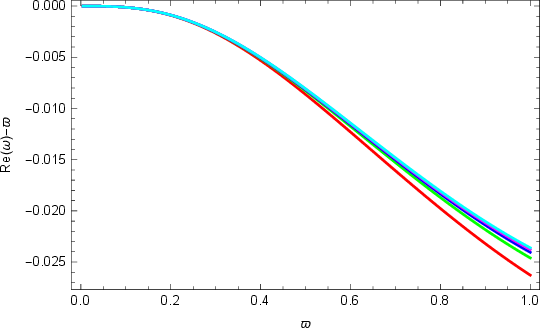}
    \includegraphics[scale=0.6]{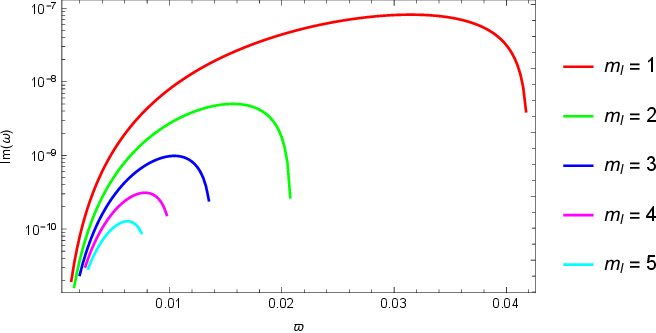}
    \caption{Profile of co-rotating quasibound states frequencies with respect to the analog scalar mass $\varpi$ with fixed $\Omega_H=0.05, 0.1, 0.3,3$ (from top row to bottom row) for various value of $m_\ell$.} \label{Bomb}
\end{figure}

In Fig.~\ref{Bomb}, we investigate the effect of the rotation parameter, $\Omega_H$, on the co-rotating ($m_\ell>0$) QBS profile. Superradiant states are cut off at $m_\ell\Omega_H$ in the case of low black hole's spin, $\Omega_H<<1$ (the first two rows). Analytically, the cutoff can be estimated by considering the zero of the second term on the left hand side of the exact energy quantization condition \eqref{exactenergy}. For the case $\Omega_H<<1$, we can approximate,
\begin{gather}
\omega -m_\ell{\Omega }_H+\frac{1}{2}\sqrt{4m_\ell^2{\Omega }^2_H-1}\approx\omega -m_\ell{\Omega }_H+\frac{i}{2}. 
\end{gather}
Therefore, the quantization condition $\eqref{exactenergy}$ is now 
\begin{align}
\frac{\varpi^2 -2{\omega }^{2}}{2\sqrt{\varpi^2 -{\omega }^{2}}}-i\left(\omega -m_\ell{\Omega }_H\right)=-\left(n+\frac{1}{2}\right) , \ \ n=1,2,3,... 
\end{align}
Clearly, there exists real solution at the cutoff frequency $\omega^{cutoff} = m_\ell \Omega_H$. Interestingly, $\omega$ has positive imaginary part when superradiant condition is satisfied $\omega < m_\ell \Omega_H$. From small spin formula, we can determine ${\Omega_H}_{crit}$ which makes $\operatorname{Im}(\omega)=0$. For instance, we find that for $\varpi=0.2$ (top row of Fig.~\ref{QBS2}) and $m_\ell=1,2,3$, ${\Omega_H}_{crit} = 0.196,0.1,0.067$, respectively. 

Conversely, for the cases with $\Omega_H>>1$, we can approximate,
\begin{gather}
\omega -m_\ell{\Omega }_H+\frac{1}{2}\sqrt{4m_\ell^2{\Omega }^2_H-1}\approx\omega.
\end{gather}
Thus, \eqref{exactenergy} can be approximated as
\begin{align}
\frac{\varpi^2 -2{\omega }^{2}}{2\sqrt{\varpi^2 -{\omega }^{2}}}-i\omega=-n\ \ , \ \ n=1,2,3,... .
\end{align}

For this case, the real solution exists if the second term is zero, $i\omega=0$. However, $\omega=0$ violates the superradiance condition $\omega^{cutoff} = m_\ell \Omega_H>0$, therefore in the large spin regime, $\Omega_H >>1$, superradiant bound state does not exist.

Moreover, empowered by the exact formula \eqref{exactenergy}, we can also compute the superradiant QBS profile with intermediate $\Omega_H$ (the third and the fourth row of Fig.~\ref{Bomb}) and find out that for intermediate spin, superradiance cutoff does not follow $\omega^{cutoff}=m_\ell\Omega_H$ any longer. 

In addition, in Fig.~\ref{Bomb}, we plot QBS profiles of co-rotating case against the rest energy $\varpi$. We observe that $\operatorname{Re}(\omega)-\varpi$ becomes more negative as the rest energy increases. This indicates that the higher the photon intensity (the larger $\varpi$), the stronger the binding energy. For the imaginary part (right column), we display the semi-log profile of the $\operatorname{Im}(\omega)$ only for the co-rotating modes ($m_\ell>0$) since all of the counter-rotating modes ($m_\ell<0$) and the spherical mode ($m_\ell=0$) are not superrandiant. At each $m_\ell$, there exists maximum $\varpi$ that allows $\operatorname{Im}(\omega)>0$. It is observed that for small $\Omega_H$, a decrease in $m_\ell$ leads to a smaller maximum value of $\varpi$. At larger $\Omega_H$, the trend is the opposite i.e., a higher $m_\ell$ corresponds to a larger maximum value of the rest energy.



\subsection{Scalar Clouds}

Scalar clouds are stationary bound states having purely real eigenfrequency $\omega_{SC}$ that are in resonance as natural multiple of the black hole's event horizon angular velocity $\Omega_H$. The configuration is mathematically described as follows \cite{Santos:2021nbf,Garc,Ric,Herdeiro:2014goa}, 
\begin{align}
\operatorname{Re}(\omega_{SC})
&= m_\ell \Omega_H < \varpi, \ \ \ m_\ell=1,2,3,... \ , \label{cond0}\\
\operatorname{Im}(\omega_{SC}) &= 0,\label{cond1}
\end{align}
and considering the relation \eqref{angmom}, one finds that the scalar cloud is a resonant mode whose frequency matches the system's optical characteristics,
\begin{equation}
    \omega_{SC}= m_\ell \frac{m\lambda}{2\pi r_H^2\sqrt{n_0n_2I_0}}=m_\ell \frac{8mr_0^2}{\pi \lambda^3}(n_0n_2 I_0)^{\frac{3}{2}}. \label{freqasrefindx}
\end{equation}

According to the results obtained via Analytical Asymptotically Matching (AAM) 
technique, the natural number $m$ is equal to the magnetic quantum number $m_\ell$ \cite{Furuhashi,Hod:2013zza,Benone:2014ssa,Huang:2016qnk}. The AAM itself, initially used by Detweiler \cite{Detweiler:1980uk} in 1979, has been proven useful to derive black holes' scalar clouds energy formulas in the following conditions,
\begin{gather}
    a <<1, \label{cond2}\\
    \omega<< 1, \label{cond3}
\end{gather}
where $a$ is the black hole's spin. The AAM matching conditions above restrict the calculation only for slowly rotating black hole and bosons having much larger Compton wavelength comparing to the rotating black hole's size. 

Moreover, recently, Mohsen Khodadi \cite{Khodadi:2022dyi} finds that the AAM's scalar clouds are not perfectly stationary, but rather very long-lived for ultralight scalar fields. This discovery motivates us to investigate the black hole's scalar cloud by using our exact bound state solution, which has no constraint on the black hole's spin and scalar mass.

Let us now consider the general quasibound states exact energy formula \eqref{exactenergy} with $\omega=\omega_{SC}=m_\ell\Omega_H$,
\begin{equation}
\frac{\varpi^2 -2\omega_{SC}^{2}}{2\sqrt{\varpi^2 -\omega_{SC}^{2}}}-\frac{i}{2}\sqrt{4\omega_{SC}^2-1}=-n. \label{ScalarCloudEnergy}
\end{equation}
The imaginary second term of \eqref{ScalarCloudEnergy} can be minimized if we consider $\omega_{SC}=\frac{1}{2}-\epsilon$. The Taylor series expansion up to $O(\epsilon)$ is obtained as follows,
\begin{equation}
 \frac{2\varpi^2 -1}{2\sqrt{4\varpi^2-1}}+\sqrt{\epsilon}=-n,
\end{equation}
and the Taylor expansion in the terms of $\epsilon$ yields,
\begin{equation}
\varpi = \sqrt{\frac{1}{2}+2n^2-n\sqrt{1+4n^2}}+\frac{8n-\frac{2 (1 + 8 n^2)}{\sqrt{1+4n^2}}}{4\sqrt{\frac{1}{2}+2n^2-n\sqrt{1+4n^2}}}\sqrt{\epsilon}+O(\epsilon).
\end{equation}
We can define an energy ratio $\gamma_\omega=\frac{\varpi}{\omega}$, where the quantity $1-\gamma_\omega^{-1}$ is the proportion of binding energy in relation to the scalar's rest energy. We perform series expansion of $\gamma_\omega$ with respect to $\epsilon$. Thus, we obtain
\begin{equation}
\gamma_\omega= \frac{\varpi}{\omega_{SC}} 
=2\sqrt{\frac{1}{2}+2n^2-n\sqrt{1+4n^2}}+\frac{8n-\frac{2 (1 + 8 n^2)}{\sqrt{1+4n^2}}}{2\sqrt{\frac{1}{2}+2n^2-n\sqrt{1+4n^2}}}\sqrt{\epsilon}+O(\epsilon).
\end{equation}

\begin{figure}[h]
    \centering
    \includegraphics[scale=0.5]{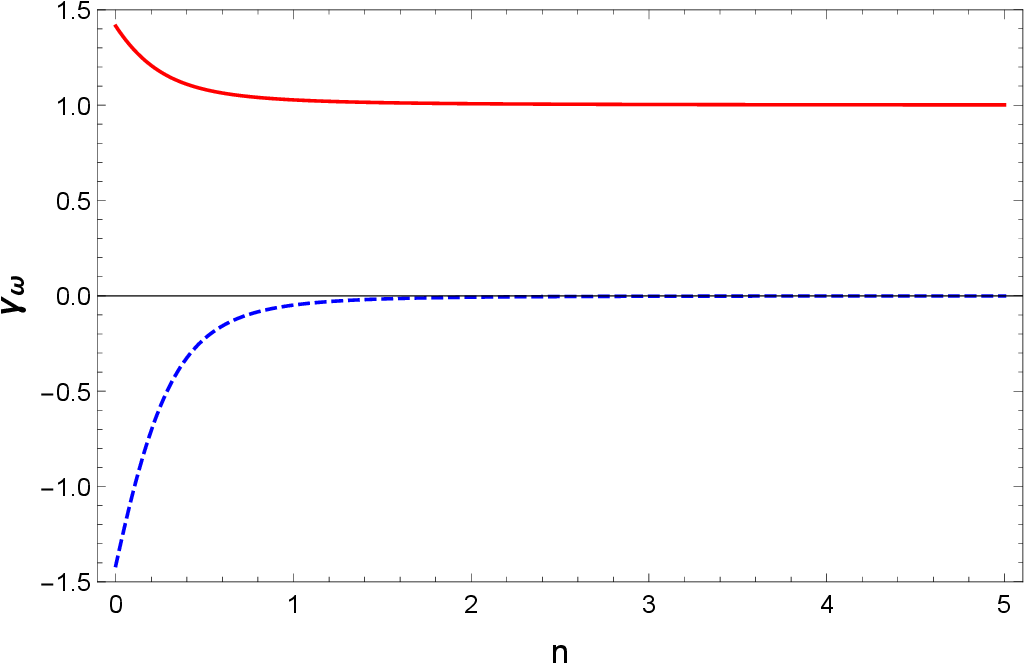}
    \caption{Profile of first (red solid line) and second terms (blue dashed line) of $\gamma_\omega$.} \label{G}
\end{figure}

In Fig.~\ref{G}, we show the behavior of the first and second terms of $\gamma_\omega$. The first term of $\gamma_\omega$ is positive definite and monotonically decreasing with respect to $n$, having a maximum at $n=1$ and saturates to unity as $n \to \infty$. The second term is negative definite and monotonically increasing with respect to $n$, having a minimum at $n=1$ and saturates to zero as $n \to \infty$.

Therefore, the upper bound of $\gamma_\omega$ is obtained from the first term with $n=1$, 
\begin{equation}
\gamma_\omega (n=1)=\sqrt{10-4\sqrt{5}}\approx1.02749\approx \sqrt{\frac{30.68}{29.06}}. \label{upperbound}
\end{equation}
This gives a more stringent constraint to $\gamma_\omega \in \left[1,1.02749\right]$ compared to the previously obtained, i.e., $\gamma_{ref} \in \left[1,\sqrt{\frac{32}{27}}\approx 1.08866\right]$, via WKB approximation by Shahar Hod \cite{Hod:2021pkw}.   In the limit $n\to\infty$,  both $\gamma_\omega$ and $\gamma_{ref}$ match, equal to unity \cite{Hod:2021qem}.

\section{Bose-Einstein Statistic on photon-fluid's Event Horizon Analog}\label{sect:BE}
In this section, the Damour-Ruffini  technique \cite{Damour} will be used to determine the boson distribution function and the Hawking temperature of the photon-fluid analog black hole's horizon. The Damour-Ruffini method utilizes the near horizon limit of the obtained exact radial solutions to find the relative probability of the outgoing wave, from where the boson distribution function is obtained. In the vicinity of the horizon, $r \to r_H$, the exact radial solutions \eqref{radialsol} have this following form,
\begin{gather}
R(r\to r_H) \approx z
\left[Az^{\frac{\beta}{2}}+Bz^{-\frac{\beta}{2}}\right], \label{wavenearh}
\end{gather}
where $\beta=2i\left(\omega -m_\ell{\Omega }_H\right).$ Remark that, we have redefined the normalization constants $A$ and $B$ to absorb the component ${\left(-1\right)}^{\frac{1}{2}(\gamma-1)}$.
 
The wave function \eqref{wavenearh} consists of two parts, i.e. the ingoing, $\psi_{+in}=A{z}^{-\frac{\beta}{2}+1}$, and outgoing, $\psi_{+out}=B{z}^{\frac{\beta}{2}+1}$, waves. When an incoming wave hits the horizon $r_H$, particle-antiparticle pairs are formed. The particle is reflected, which enhances the outgoing wave, while the antiparticle counterpart forms the transmitted wave that crosses the horizon absorbed by the black hole. Analytical continuation of the wave function $\psi\left(z\right)$ may be determined using the following trick,
\begin{equation}
{z}^{\lambda }\to {\left[\left(\frac{r}{r_H }-1\right)+i\epsilon \right]}^{\lambda } =\left\{ \begin{array}{cc}
{z}^{\lambda } &,\ r>r_H  \\ 
{\left\lvert z\right\rvert }^{\lambda }e^{i\lambda \pi} &,\ r<r_H  \end{array}
\right..
\end{equation}
This allows us to obtain $\psi_{-out}$ from $\psi_{+out}$ as follows,
\begin{gather}
    \psi_{-out}=\psi_{+out}(  z\to ze^{i\pi}).
\end{gather}
Explicitly, we obtain this following relation,
\begin{equation}
\begin{split}
\psi_{-out}&=A{\left(ze^{i\pi}\right)}^{\frac{\beta}{2}+1},\\
&=\psi_{+out}e^{\frac{i\pi \beta}{2}},
\end{split}
\end{equation}
where,
\begin{equation}
{\left\lvert \frac{\psi_{+out}}{\psi_{+in}}\right\rvert }^2={\left\lvert \frac{\psi_{-out}}{\psi_{+in}}\right\rvert }^2e^{-2\pi\beta i}={\left\lvert \frac{\psi_{-out}}{\psi_{+in}}\right\rvert }^2e^{4\pi\left(\omega-m_\ell\Omega_H\right)}  .  
\end{equation}
With the help of the step function $\theta(z)$, we can rewrite the outgoing waves as follows,
\begin{align}
\psi_{out}&=\psi_{-out}\theta(-z)+\psi_{+out}\theta(z)\nonumber\\&=A{z}^{\frac{\beta}{2}+1}\left[e^{2\pi\left(\omega-m_\ell\Omega_H\right)}\theta \left(-z\right)+\theta \left(z\right)\right].
\end{align}
These wave functions are subjected to the following normalization condition, 
\begin{align}
\left\langle \frac{\psi_{out}}{\psi_{in}}\mathrel{\left\lvert \vphantom{\frac{\psi_{out}}{\psi_{in}} \frac{\psi_{out}}{\psi_{in}}}\right.\kern-\nulldelimiterspace}\frac{\psi_{out}}{\psi_{in}}\right\rangle &= 1 = {\left\lvert \frac{A}{B}\right\rvert }^2\left\lvert 1-e^{4\pi\left(\omega-\Omega_H\right)}\right\rvert, \\
\left\lvert \frac{A}{B}\right\rvert ^2 &= \frac{1}{e^{4\pi\left(\omega-m_\ell\Omega_H\right)}-1}.
\end{align}
Note that the modulus square of the ratio of the outgoing to the incoming waves yields a Bose-Einsteinian radiation distribution function that can be modified as follows, 
\begin{gather}
4\pi\left(\omega-m_\ell\Omega_H\right)=4\pi\left(\frac{Er_H}{\hbar c}-m_\ell\Omega_H\right)=\frac{\hslash \left(\omega -\mu_H\right)}{\left(\frac{c\hbar}{4\pi r_H}\right)},
\end{gather} 
where $\mu_H=\frac{c m_\ell\Omega_H}{r_H}$ is the gravitational analog chemical potential. Lastly, we determine the analog black hole's horizon temperature by comparing it with the Bose-Einstein distribution function $e^{\frac{\hbar(\omega-\mu)}{k_B T}}$ as follows, 
\begin{gather}
T=T_H=\frac{c\hbar}{4\pi k_B r_H}=I_0r_0\frac{n_0n_2c\hbar}{\pi k_B \lambda^2},\label{hawking}
\end{gather} 
where $T_H$ is the Hawking temperature. Interestingly, the Hawking temperature of the photon-fluid black hole analog has the same form with the chargeless Lense-Thirring black hole \cite{senjaya1}, $T_H=\frac{r_Hc\hslash }{4\pi k_Br^2_s}$, where $r_s=\frac{2GM}{c^2}$ is the Schwarzschild radius.

Additionally, the energy flow by Hawking radiation might be further calculated analytically as follows,
\begin{align}
\Phi_E &= \int_0^\infty \frac{\omega}{e^{4\pi\left(\omega-m_\ell\Omega_H\right)}-1} d\omega, \\
&= \frac{1}{16\pi^2}Li_2\left(e^{4\pi m_\ell\Omega_H}\right), \label{hawkingflux}
\end{align}
where the function $Li_n(x)$ is the Jonqui\'ere's Polylogarithm function \cite{NIST}. 

\begin{figure}[h]
    \centering
    \includegraphics[scale=1]{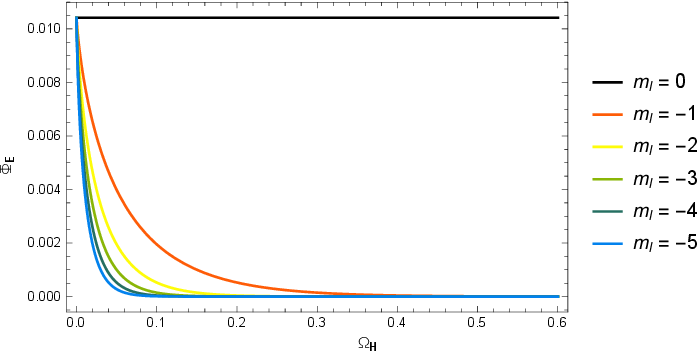}
    \caption{Hawking radiation flux profile as functions of black hole's angular momentum parameter, $\Omega_H$, for various $m_\ell$.} 
  \label{hawking1}
\end{figure}

The Hawking radiation flux, $\Phi_E$, is displayed in \cref{hawking1} as a function of the black hole's angular momentum parameter, $\Omega_H$, for a range of $m_\ell$. The bosonic radiation flux has a constant value for $m_\ell=0$. For $m_\ell\neq 0$, the radiation flux is maximum at the small angular momentum $\Omega_H$ and approaches zero at large $\Omega_H$. Furthermore, it should be noted that positive $m_\ell$ is complex-valued and, therefore, omitted. 

\section{Superradiance and Wave Amplification}\label{sect:suprad}

A spinning geometry is necessary for superradiance, and in the photon-fluid model, this is represented by the angular momentum, $\Omega_H$. Analytical investigation into black hole superradiance can be done by investigating the so called amplification factor, i.e. the modulus square of the ratio of the outgoing scattered wave to the ingoing wave. Positive amplification factor signifies the presence of superradiance, whereas negative amplification denotes the absence of superradiance.

In this section, we will derive the wave amplification coefficient of scalar fields around the analog spinning black hole using the AAM technique. The following describes the AAM procedure: Finding analytical solutions at two extremes-near the event horizon and at infinity-is the first step. The far field limit of the near the event horizon solution and the near horizon limit of the far field solution are analytically constructed by utilizing connection formulae of the hypergeometric functions. We match both solutions in the so-called intermediate region. After obtaining the matching coefficients, the amplification factor can be obtained. 

Let us start with the Klein-Gordon equation in photon-fluid black hole spacetime \eqref{radialz}. The radial variable $z=-\frac{r-r_H}{r_H}$ has domain $-\infty<z\leq 0$ and we will transform $z\to-z$ to shift the domain to $0\leq z <\infty$. After changing the sign of $z$ and multiplying the radial equation \eqref{radialz} by $z^2(z+1)^2$, we obtain,
\begin{align}
0 &= z^2(z+1)^2{\partial }^{2}_zR+P\left(z\right){\partial }_zR+Q\left(z\right)R, \label{modrad}
\end{align}
where
\begin{align}
P\left(z\right) &= z(z+1), \label{modP}\\
Q\left(z\right) &= {\left(\omega -m_\ell{\omega }_H\right)}^2 +\left(-\frac{1}{2}-m_\ell^2 + 4 \omega^2 -\varpi^2 - 4 m_\ell \omega \Omega_H\right)z \nonumber \\
&~~~~+\left(\frac{1}{4}-  m_\ell^2 + 6 \omega^2 - 3\varpi^2 - 
 2 m_\ell \omega \Omega_H \right)z^2\nonumber \\
 &~~~~+\left(4 \omega^2-3 \varpi^2 \right)z^3+\left( \omega^2-\varpi^2 \right)z^4. \label{modQ}
\end{align}

\subsection{Far Field Solution}
Let us consider the radial equation \eqref{modrad} and take the limit $z\to\infty$, we find
\begin{gather}
{\partial }^{2}_zR+\left[A_0+\frac{A_1}{z}+\frac{A_2}{z^2}\right] R=0, \label{normalformfarfieldequation}\\
A_0=\omega^2-\varpi^2,\\
A_1=4\omega^2- 3\varpi^2 ,\\
A_2=\frac{1}{4}- 
 m_\ell^2 + 6 \omega^2 - 3 \varpi^2 - 2 \omega m_\ell\Omega_H. \label{A2far}
\end{gather}
After rescaling $\Tilde{z} \to i\alpha z$, \eqref{normalformfarfieldequation} can be put into the Whittaker equation (see Appendix \ref{AppendixD}). The general solutions are then written in the confluent hypergeometric functions $F_{1,1}\left(a,b,x\right)$ as follows 
\begin{multline}
   R_{far}=N_{\infty 1}e^{-i\frac{\alpha}{2} z} z^{\frac{1}{2}+\beta} F_{1,1}\left(\frac{1}{2}+\beta+i\frac{A_1}{\alpha},1+2\beta,i\alpha z\right) \\+ N_{\infty 2}e^{-i\frac{\alpha}{2} z} z^{\frac{1}{2}-\beta}F_{1,1}\left(\frac{1}{2}-\beta+i\frac{A_1}{\alpha},1-2\beta,i\alpha z\right), \label{farsol}
\end{multline}
where $N_{\infty 1}$ and $N_{\infty 2}$ are arbitrary constants and,
\begin{gather}
    \alpha=2\sqrt{A_0}, \ \ \  \ \ \ \beta=\frac{1}{2}\sqrt{1-4A_2}.
\end{gather}
The near horizon limit of the far solution is given by, 
\begin{equation}
R_{far}\left(z\to 0\right)\approx N_{\infty 1}z^{\frac{1}{2}+\beta}+N_{\infty 2}z^{\frac{1}{2}-\beta}.
\end{equation}
\subsection{Near Horizon Solution}
The next step is to determine the solutions to the radial equation \eqref{modrad}  near the event horizon of the black hole, $z\to 0$. Taking the limit $z\to 0$, we obtain,
\begin{gather}
z^2(z+1)^2{\partial }^{2}_zR+z(1+z)\partial_zR+\left(B_0+{B_1}z+B_2z^2\right)R=0, \label{10}\\
B_0=(\omega- m_\ell \Omega_H )^2 ,\\
B_1=-\frac{1}{2}- 
 m_\ell^2 + 4 \omega^2 - \varpi^2  \textcolor{blue}{-} 4 \omega m_\ell  \Omega_H,\\
B_2=\frac{1}{4}- 
 m_\ell^2 + 6 \omega^2 - 3 \varpi^2 \textcolor{blue}{-} 2 \omega m_\ell\Omega_H. \label{B2near}
\end{gather}

Comparing the near horizon equation \eqref{10} with Gauss hypergeometric function \eqref{gauss1} of Appendix \ref{AppendixE}, we obtain the following near horizon solution that represents ingoing wave,
\begin{equation}
R_{near}=N_{H1}(z+1)^{1-\sqrt{1-B_0+B_1-B_2}} z^{-i(\omega- m_\ell \Omega_H)}F_{2,1}\left(a_1,a_2,a_3,-z\right),
\end{equation}
where,
\begin{gather}
    a_1=\frac{1}{2}\left(1+\sqrt{1-4B_2}\right)-i(\omega- m_\ell \Omega_H)-\sqrt{1-B_0+B_1-B_2},\\
    a_2=\frac{1}{2}\left(1-\sqrt{1-4B_2}\right)-i(\omega- m_\ell \Omega_H)-\sqrt{1-B_0+B_1-B_2},\\
    a_3=1-2i(\omega- m_\ell \Omega_H).
\end{gather}
We notice that $
1-B_0+B_1-B_2=\frac{1}{4} - 3 \omega^2 + 2 \varpi^2 - m_\ell^2 \Omega_H^2.
$ Now the far field limit of the near horizon solution can be derived using the identity \eqref{connection} as follows, 
\begin{multline}
   R_{near}\left(z\to \infty \right)\approx N_{H1}z^{-\left(-1+\sqrt{1-B_0+B_1-B_2}+i(\omega- m_\ell \Omega_H)\right)}\times\\ \left[z^{-a_1}\frac{\Gamma \left(a_3\right)\Gamma \left(a_2-a_1\right)}{\Gamma \left(a_3-a_1\right)\Gamma \left(a_2\right)}+z^{-a_2}\frac{\Gamma \left(a_3\right)\Gamma \left(a_1-a_2\right)}{\Gamma \left(a_3-a_2\right)\Gamma \left(a_1\right)}\right]. 
\end{multline}

We can rearrange the terms as follows,
\begin{multline}
R_{near}\left(z\to \infty \right)\approx N_{H1}   \frac{\Gamma \left(a_3\right)\Gamma \left(a_2-a_1\right)}{\Gamma \left(a_3-a_1\right)\Gamma \left(a_2\right)}z^{-\left(-1+\sqrt{1-B_0+B_1-B_2}+i(\omega- m_\ell \Omega_H)\right)-a_1}\\+N_{H1} \frac{\Gamma \left(a_3\right)\Gamma \left(a_1-a_2\right)}{\Gamma \left(a_3-a_2\right)\Gamma \left(a_1\right)}z^{-\left(-1+\sqrt{1-B_0+B_1-B_2}+i(\omega- m_\ell \Omega_H)\right)-a_2}.
\end{multline}

\subsection{Matching The Solutions}
Further investigation into the exponents of $z$ reveals that,
\begin{gather}
-\left(-1+\sqrt{1-B_0+B_1-B_2}+i(\omega- m_\ell \Omega_H)\right)-a_1=\frac{1}{2}-\frac{1}{2}\sqrt{1-4B_2},\\
-\left(-1+\sqrt{1-B_0+B_1-B_2}+i(\omega- m_\ell \Omega_H)\right)-a_2=\frac{1}{2}+\frac{1}{2}\sqrt{1-4B_2}.
\end{gather}
This tells us that in the overlap region, we can match $R_{far}\left(z\to 0\right)$ and $R_{near}\left(z\to \infty \right)$ if this following condition is fulfilled,
\begin{equation}
A_2 - B_2 =0.
\end{equation}
From \eqref{A2far} and \eqref{B2near}, these confirms the matching condition.
The following relationships are obtained if the AAM matching requirement above is met,
\begin{gather}
N_{\infty 1}=N_{H1}\frac{\Gamma \left(a_3\right)\Gamma \left(a_1-a_2\right)}{\Gamma \left(a_3-a_2\right)\Gamma \left(a_1\right)}, \label{match1}\\
N_{\infty 2}=N_{H1}\frac{\Gamma \left(a_3\right)\Gamma \left(a_2-a_1\right)}{\Gamma \left(a_3-a_1\right)\Gamma \left(a_2\right)}. \label{match2}
\end{gather}
Now we shall investigate the far field solutions \eqref{farsol} in their asymptotic far limit. As $z\to\infty$, the confluent hypergemetric function $F_{1,1}$ can be transformed using the rule stated in \eqref{faratfar}. Let us begin with \eqref{farsol}'s first term as follows,
\begin{multline}
N_{\infty 1}e^{\frac{1}{2}i\alpha z}z^{\frac{1}{2}+\beta}F_{1,1}\left(\frac{1}{2}+\beta+i\frac{A_1}{\alpha}, 1+2\beta,\ i\alpha z\right) =  \\
N_{\infty 1}\Bigg[(i\alpha)^{-\frac{1}{2}-\beta+i\frac{A_1}{\alpha}}z^{i\frac{A_1}{\alpha}}e^{i\frac{\alpha}{2}z}\frac{\Gamma\left(1+2\beta\right)}{\Gamma\left(\frac{1}{2}+\beta+i\frac{A_1}{\alpha}\right)}  \\   
+ (-i\alpha)^{-\frac{1}{2}-\beta-i\frac{A_1}{\alpha}}z^{-i\frac{A_1}{\alpha}}e^{-i\frac{\alpha}{2}z}\frac{\Gamma\left(1+2\beta\right)}{\Gamma\left(\frac{1}{2}+\beta-i\frac{A_1}{\alpha}\right)} \Bigg],
\end{multline}
while the expansion of the second part gives,
\begin{multline}
N_{\infty 2}e^{-\frac{1}{2}i\alpha z}z^{\frac{1}{2}-\beta}F_{1,1}\left(\frac{1}{2}-\beta+i\frac{A_1}{\alpha},\ 1-2\beta,\ i\alpha z\right)=\\
    N_{\infty 2}\Bigg[(i\alpha)^{-\frac{1}{2}+\beta+i\frac{A_1}{\alpha}}z^{i\frac{A_1}{\alpha}}e^{i\frac{\alpha}{2}z}\frac{\Gamma\left(1-2\beta\right)}{\Gamma\left(\frac{1}{2}-\beta+i\frac{A_1}{\alpha}\right)} 
    \\+(-i\alpha)^{-\frac{1}{2}+\beta-i\frac{A_1}{\alpha}} z^{-i\frac{A_1}{\alpha}}e^{-i\frac{\alpha}{2}z}\frac{\Gamma\left(1-2\beta\right)}{\Gamma\left(\frac{1}{2}-\beta-i\frac{A_1}{\alpha}\right)}\Bigg].
\end{multline} 
As a superposition of ingoing and outgoing waves, the asymptotically distant, far field solutions are ultimately obtained as follows,
\begin{align}
  R_{far}(z\to \infty)&= A_{in} e^{-\frac{1}{2}i\alpha z} z^{-i\frac{A_1}{\alpha}} +A_{out} e^{\frac{1}{2}i\alpha z} z^{i\frac{A_1}{\alpha}},
\end{align}
with,
\begin{align}
A_{out} &\equiv N_{\infty 1}(i\alpha)^{-\frac{1}{2}-\beta+i\frac{A_1}{\alpha}}\frac{\Gamma\left(1+2\beta\right)}{\Gamma\left(\frac{1}{2}+\beta+i\frac{A_1}{\alpha}\right)}+N_{\infty 2}(i\alpha)^{-\frac{1}{2}+\beta+i\frac{A_1}{\alpha}}\frac{\Gamma\left(1-2\beta\right)}{\Gamma\left(\frac{1}{2}-\beta+i\frac{A_1}{\alpha}\right)},\label{Ain}\\
A_{in}&\equiv N_{\infty 1}(-i\alpha)^{-\frac{1}{2}-\beta-i\frac{A_1}{\alpha}}\frac{\Gamma\left(1+2\beta\right)}{\Gamma\left(\frac{1}{2}+\beta-i\frac{A_1}{\alpha}\right)}+N_{\infty 2}(-i\alpha)^{-\frac{1}{2}+\beta-i\frac{A_1}{\alpha}}\frac{\Gamma\left(1-2\beta\right)}{\Gamma\left(\frac{1}{2}-\beta-i\frac{A_1}{\alpha}\right)}. \label{Aout} 
\end{align}
Remark that, the constants $N_{\infty1,2}$ can be written in term of the near horizon constant $N_{H1}$ via \eqref{match1} and \eqref{match2}. The amplification factor is defined as follows \cite{refcoef,Suphot,Brito},
\begin{align}
    Z_{m_\ell,\Omega_H }&\equiv {\left|\frac{A_{out}}{A_{in}}\right|}^2-1.\label{SuperAmp}
\end{align}
This has a physical meaning as follows: 
\begin{itemize}
    \item $ Z_{m_\ell,\Omega_H }>0$ implies a superradiance and a gain in amplification factor.
    \item $ Z_{m_\ell,\Omega_H }<0$ indicates a loss in amplification, which corresponds to the absence of the superradiance.
    \item Superradiance threshold, $\omega=\Omega_H $ (see \eqref{cond0}), indicates the point at which $Z_{m_\ell,\Omega_H }(\Omega_H )=0$.
\end{itemize}

Additionally, one should consider $\{\omega,\varpi\}<< 1$ condition, which implies $\frac{1}{\omega}>>1$. In the region $\omega z <<1$, we can approximate \cite{xo,st},
\begin{align}
B_0 &= (\omega- m_\ell \Omega_H )^2 ,\\
B_1 &\approx -\frac{1}{2}- 
 m_\ell^2 - 4 \omega m_\ell  \Omega_H,\\
B_2 &\approx \frac{1}{4}- 
 m_\ell^2 - 2 \omega m_\ell\Omega_H.
\end{align}
Additionally, in the region $ z>>\frac{1}{4}-m_\ell^2>1$, one obtains,
\begin{align}
A_0 &= \omega^2-\varpi^2,\\
A_1 &\approx 0 ,\\
A_2 &\approx \frac{1}{4}- 
 m_\ell^2 - 2 \omega m_\ell\Omega_H.
\end{align}
It is crucial to remember that the scalar energy has an impact on the width of the matching region, which contains the AAM solution. We can run the following analysis, 
\begin{enumerate}
    \item In the region nearby the event horizon, the condition $\omega z <<1$ is equivalent to,
    \begin{equation}
        r-r_H<<\frac{r_H}{\omega}. \label{111}
    \end{equation}
    \item In the far region, the requirement $z>>1$ is equal to,
    \begin{equation}
        r_H<<r-r_H. \label{112}
    \end{equation}
\end{enumerate}
Combining \cref{111,112} yields the matched region,
\begin{equation}
    r_H<<r-r_H<<\frac{r_H}{\omega}. \label{AAMregion}
\end{equation}
Therefore, the smaller $\omega$, the wider the matching zone, resulting in better approximation.

\begin{figure}[h]
    \centering
    \includegraphics[scale=0.9]{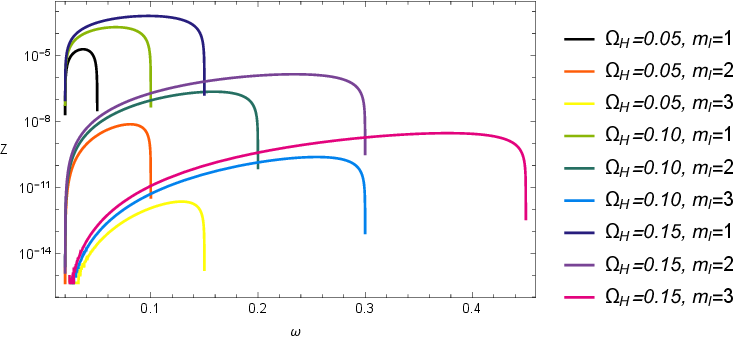}
    \caption{Amplification factor profile for various combinations of $m_\ell$ and $\Omega_H$ with fixed $\varpi=0.02$. } 
  \label{Amplification}
\end{figure}

In Fig. \ref{Amplification}, we examine how the amplification factor $Z_{m_\ell,\Omega_H }$ \eqref{SuperAmp} behaves in relation to $\omega$. The only cases with positive amplification factors are those with $m_\ell>0$. We observe that increasing the angular momentum parameter $\Omega_H $ raises the maximum value of $Z_{m_\ell,\Omega_H }$ for the given value of $m_\ell$. One can also identify the superradiant threshold, i.e. when $Z_{m_\ell,\Omega_H }$ quenches, occurs at $\omega=m_\ell \Omega_H $. Therefore, as $m_l$ increases, we find that the amplification factor remains positive for wider range of $\omega$. We notice that configuration with higher $m_l$ has smaller amplification factor. This means that the lowest $m_l=1$ mode is amplified the most. The 3+1 dimension Kerr family black holes amplification factor exhibits similar behavior with the 2+1 dimension photon-fluid analog black hole's \cite{Brito,Franzin,Senjaya:2025bbp}.

\begin{figure}[H]
    \centering
    \includegraphics[scale=0.9]{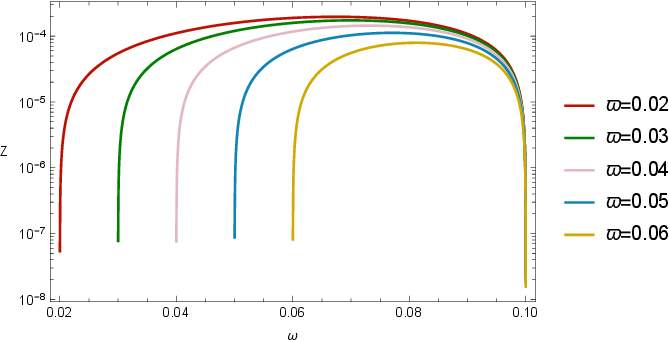}
    \caption{Amplification factor profile  for various $\varpi$ with fixed $\Omega_H =0.1, \ m_\ell=1$. } 
  \label{Amplification1}
\end{figure}

In Fig. \ref{Amplification1}, we examine the impact of scalar field mass on the black hole's amplification factor, setting $\Omega_H =0.1$ and $m_\ell=1$ for all cases. Scalar fields with lighter masses have a wider frequency range in the superradiant domain since superradiance condition ranges between $\varpi$ and $m_\ell \Omega_H $. Furthermore, it should be noted that the cutoff frequency is not influenced by the scalar mass but rather by the characteristics of the black hole.

To summarize this subsection, we find that a photon fluid system  exhibits superradiant energy extraction in the same way that astrophysical black holes theoretically do. A photon fluid system, which is a nonlinear optical medium in which light behaves as a collective fluid, can mimic rotating curved spacetime geometries \cite{Braidotti:2020ize}. When incident electromagnetic excitations with energy $\omega$ in the range $\varpi<\omega < m_\ell \Omega_H$ come, the waves are scattered off by the rotating analogue black hole and amplified, 
extracting the black hole's rotational energy.

\section{Greybody Factor}\label{sect:gbf}

\begin{figure}[h]
    \centering
        \includegraphics[scale=0.5]{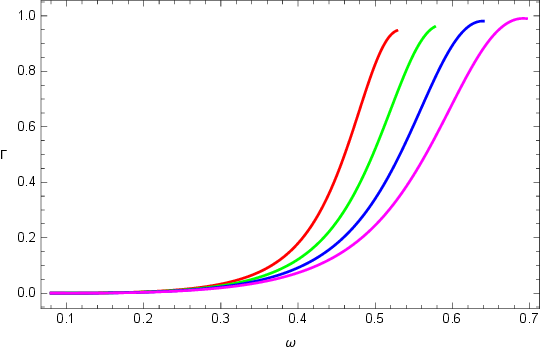}
         \includegraphics[scale=0.535]{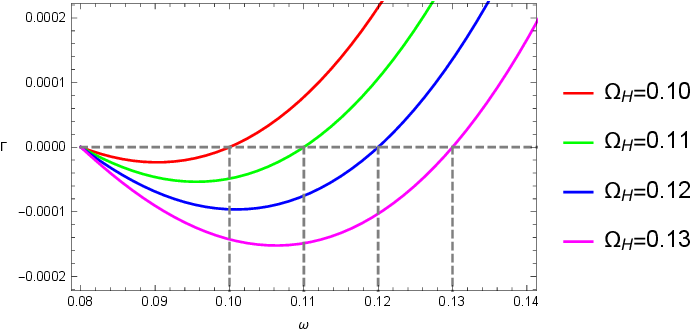}
    \caption{Greybody factor profile  for various $\Omega_H$ with fixed $\varpi=0.08, m_\ell=1$. } 
  \label{coRGBS1}
\end{figure}

\begin{figure}[h]
    \centering
        \includegraphics[scale=0.5]{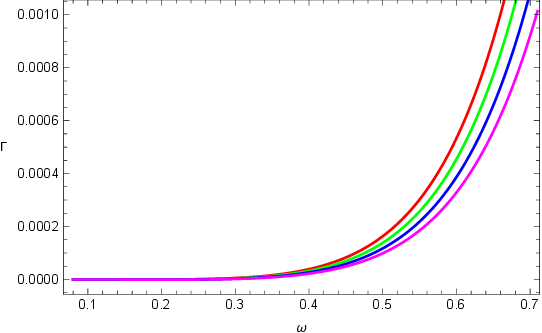}
         \includegraphics[scale=0.52]{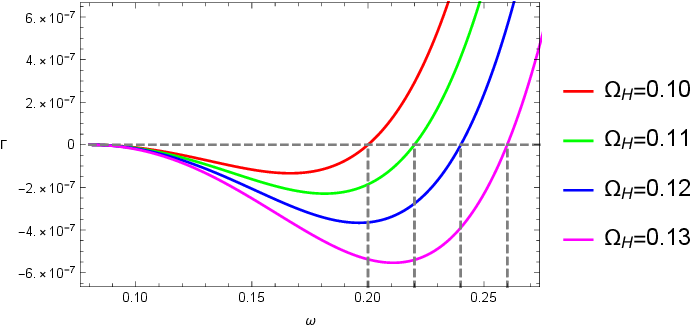}
    \caption{Greybody factor profile  for various $\Omega_H$ with fixed $\varpi=0.08, m_\ell=2$. } 
  \label{coRGBS2}
\end{figure}

\begin{figure}[h]
    \centering
        \includegraphics[scale=0.52]{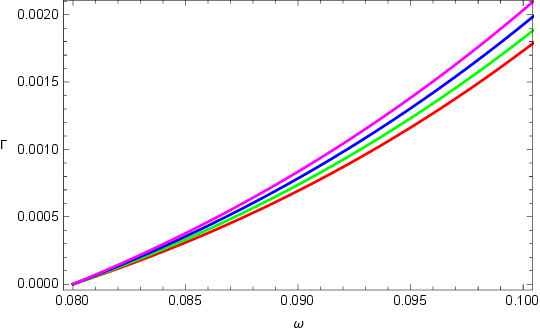}
        \includegraphics[scale=0.5]{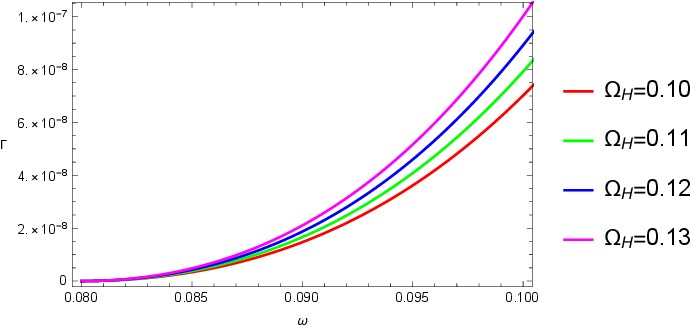}
    \caption{Greybody factor profile  for various $\Omega_H$ with fixed $\varpi=0.08, m_\ell=-1$ (left) and $m_\ell=-2$ (right). } 
  \label{CounRGBS1}
\end{figure}

Unlike black-body radiation from a black box, black hole can radiate black-body spectrum that are dependent on black hole's gravitational effective potential, which has to be tunnelled through by the emitted radiation beam. As a consequence, a distant observer will see a spectrum different from the original beam. The frequency dependent tunnelling probability is known as the greybody factor $\Gamma_{m_\ell,\Omega_H }$ \cite{Sharif:2022yxg,Javed:2024jty}. The greybody factor describes the probability
for an outgoing wave, with energy $\omega$, to reach infinity that coincides with the absorption probability for an incoming wave, with energy $\omega$, to reach the black hole's horizon \cite{Harmark:2007jy,Hyun:2019qgq}
\begin{equation}
 \Gamma_{m_\ell,\Omega_H }\equiv 1-{\left|\frac{A_{out}}{A_{in}}\right|}^2.
\end{equation}
Therefore, the greybody factor can be determined from \eqref{Ain} and \eqref{Aout}. Here, we plot the behavior of the greybody factor for low frequency $\omega$ in Fig.~\ref{coRGBS1}--\ref{CounRGBS1}. We fix $\varpi=0.08$ and $m_\ell=+1,+2$ and $m_\ell=-1,-2$, respectively, in the photon-fluid black hole analog spacetime with respect to $\omega$ for various $\Omega_H$. 

In the co-rotating mode, $m_\ell\geq 1$, there are regions with negative $\Gamma$, i.e. $\varpi < \omega < m_\ell \Omega_H$, where superradiant occurs causing the amplitude of the reflected wave exceeds that of the incident wave (see Fig.~\ref{coRGBS1} and \ref{coRGBS2}). The right plots display a close up $\Gamma$ at very small $\omega$. These show the negative region of $\Gamma$. We also mark the location of superradiant threshold  $\omega=m_\ell \Omega_H$ which clearly indicated by $\Gamma=0$. As black hole rotates faster i.e., higher $\Omega_H$, the greybody factor becomes more negative at a given $\omega$. Similar trends are also observed for $m_\ell=2$. However, the magnitude of negative $\Gamma$ is significantly smaller than those of $m_\ell=1$ case. This indicates that the most fundamental co-rotating mode experiences the most superradiant effect. 

In contrast, the counter-rotating mode, $m_\ell\leq0$, $\Gamma$ is consistently positive for all $\omega$ indicating that there is no superradiance as can be seen in Fig.~\ref{CounRGBS1}. At a given $\omega$, the greybody factor $\Gamma$ of higher spin is greater than the lower spin case.

\begin{figure}[h]
    \centering
        \includegraphics[scale=0.52]{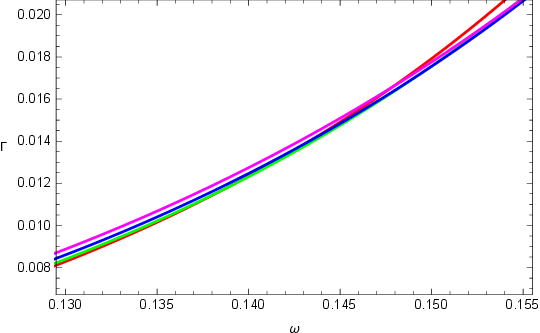}
        \includegraphics[scale=0.5]{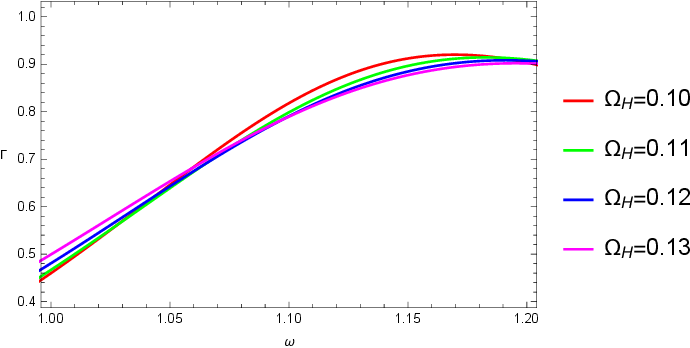}
    \caption{The crossing point of the Greybody factor for various $\Omega_H$ with fixed $\varpi=0.08, m_\ell=-1$ (left) and $m_\ell=-2$ (right). } 
  \label{CounRGBS2}
\end{figure}


\section{Conclusions}\label{sect:conclud}
In this work, we investigate the spectroscopy of 2+1 dimensional analog rotating black hole in photon-fluid model. In the sound wave regime, the dynamics of acoustic density fluctuations in the presence of vortex flows is described by an analog massive Klein-Gordon equation in a non-trivial 2+1-dimensional curved spacetime with radial wave equation given by \eqref{mainradeq}. The exact solution to the analog massive Klein-Gordon equation are derived and presented in terms of the Confluent Heun functions \eqref{radialsol}. 

We move on to investigate the analog quasibound states, which are specified by boundary conditions: purely ingoing near the black hole's event horizon and disappearing far away from it. The radial solution's polynomial condition yields the quasibound state eigenenergies, which are shown in \eqref{exactenergy}. We find that co-rotating quasibound states ($m_\ell>0$) may exhibit superradiant instabilities. However, the instability is vanishing for large black hole spin. The upper threshold of superradiant condition defines scalar clouds, which are stationary bound states with purely real eigenfrequency. With exact quantization formula in hand, we find a more stringent constraint to the scalar cloud energy ratio $\gamma_\omega\approx1.02749$, revisiting the previously obtained via WKB approximation $\gamma_\omega\approx1.08866$ \cite{Hod:2021pkw}.

In the neighborhood of the analog black hole's horizon, we perform linear series expansion of our exact radial solutions, which represent ingoing and outgoing waves. We implement the Damour-Ruffini approach to calculate the modulus square of the ratio of outgoing to incoming waves. This yields a Bose-Einsteinian distribution function, which is used to extract the Hawking temperature of the analog black hole's horizon. Additionally, we also present analytical expression of the Hawking radiation flux by integrating the distribution function, resulting in the Jonquiere’s Polylogarithm function \eqref{hawkingflux}. 

The presence of superradiant effect leads us to explore amplification factor. We use the AAM to derive analytical expression of the scattering amplification factor \eqref{SuperAmp}. The analog spinning black hole amplifies all co-rotating incoming waves in the energy range $\varpi<\omega<m_\ell \Omega_H$. In regards to the black hole's angular momentum, the amplification factor increases with the value of the black hole's spin. 

Finally, we examine the greybody factor of the analog black hole, which indicates the probability of Hawking radiation reaching a distant observer. In the low energy domain, we employ the AAM approach to compute the greybody factor. In the superradiant regime, where $\varpi < \omega < m_\ell \Omega_H$, only co-rotating modes experience negative greybody factors. For counterrotating modes, all greybody factors are positive, indicating the absence of superradiance.

The next phase in this research is to expand the investigation into the following two important directions, the analog black hole shadow, to study how the effective geometry of a photon fluid black hole analog produces an observable shadow, drawing parallels with gravitational black hole imaging and the analog black hole thermodynamics, i.e. investigating thermodynamic variables such as surface gravity, horizon temperature, and entropy using the photon fluid framework to gain a better understanding of horizon physics in the analogue system.

\section*{Acknowledgments}
SP acknowledges funding support from the NSRF via the Program Management Unit for Human Resources \& Institutional Development, Research and Innovation [grant number B39G680009].

\begin{appendices} 
    \include{appendix} 
\section{Derivation to The Analog Klein-Gordon Equation in Photon-Fluid Model} \label{Appendix0}
The analog matric tensor \eqref{metric} can be written in a matrix form as follows,
\begin{equation}
g_{\mu \nu }=\left( \begin{array}{ccc}
g_{00} & 0 & g_{02} \\ 
0 & g_{11} & 0 \\ 
g_{20} & 0 & g_{22} \end{array}
\right),
\end{equation}
where $(0,1,2)$ stands for the coordinates $(ct,r,\theta)$ and the matrix inverse is given by,
\begin{equation}
g^{\mu \nu }=\frac{1}{g}\left( \begin{array}{ccc}
g_{11}g_{22} & 0 & -g_{02}g_{11} \\ 
0 & g_{00}g_{22}-g^2_{02} & 0 \\ 
-g_{02}g_{11} & 0 & g_{00}g_{11} \end{array}
\right),    
\end{equation}
where $g$ is the metric tensor determinant,
\begin{equation}
    g=g_{11}\left(g_{00}g_{22}-g^2_{02}\right)=-r^2.
\end{equation}

Let us work out the explicit form of the Laplace-Beltrami operator in \eqref{KGEq} component by component with $\psi$ given by the separation ansatz \eqref{ansatz},
\begin{gather}
\frac{1}{\sqrt{-g}}{\partial }_0\left(\sqrt{-g}g^{00}{\partial }_0\right)\psi=\frac{1}{\Delta}\left(\frac{E}{\hbar c}\right)^2\psi,\\
\frac{1}{\sqrt{-g}}{\partial }_0\left(\sqrt{-g}g^{02}{\partial }_2\right)\psi=\frac{1}{\sqrt{-g}}{\partial }_2\left(\sqrt{-g}g^{20}{\partial }_0\right)\psi=-\frac{m_\ell \omega_H r_H^2}{\Delta r^2}\frac{E}{\hbar c}\psi,\\
\frac{1}{\sqrt{-g}}{\partial }_2\left(\sqrt{-g}g^{22}{\partial }_2\right)\psi=-\frac{m_\ell^2}{\Delta r^2}\left(\Delta-\frac{\omega_H^2r_H^4}{r^2}\right)\psi,\\
\frac{1}{\sqrt{-g}}{\partial }_1\left(\sqrt{-g}g^{11}{\partial }_1\right)\psi=\frac{e^{-i\frac{E}{\hbar c}ct}e^{im_\ell\theta}}{\Delta\sqrt{r}}\left[\Delta\partial_r(\Delta\partial_r)+\Delta\left(-\frac{\partial_r \Delta}{2r}+\frac{{\Delta}}{4r^2}\right)\right]R.
\end{gather}

Combining the first three components of the Laplace-Beltrami operator, we obtain,
\begin{equation}
\frac{1}{\Delta}\left(\frac{E}{\hbar c}-\frac{m_\ell\omega_H r_H^2}{r^2}\right)^2-\frac{m_\ell^2}{r^2},    
\end{equation}
and after the substitution of the last part into \eqref{KGEq}, we arrive at this following expression,
\begin{multline}
 \left[\frac{1}{\Delta}\left(\frac{E}{\hbar c}-\frac{m_\ell\omega_H r_H^2}{r^2}\right)^2-\frac{m_\ell^2}{r^2}\right]R+\\ \frac{1}{\Delta}\left[\Delta\partial_r(\Delta\partial_r)+\Delta\left(-\frac{r_H}{2r^3}+\frac{{\Delta}}{4r^2}\right)\right]R-\frac{\varpi^2}{\hbar^2}R=0.
\end{multline}

Multiplying the whole equation with $\Delta$ and slightly rearranging the terms yields,
\begin{equation}
\Delta {\partial }_r\left(\Delta {\partial }_r\right)R+\left[\Delta \left(\frac{\Delta }{4r^2}-\frac{r_H}{2r^3}-\frac{m_\ell^2}{r^2}-{\omega }^2_0\right)+{\left(\frac{E}{\hbar c} -\frac{m_\ell{\omega }_H r_H^2}{r^2}\right)}^2\right]R=0.  \label{maineqGBF}   
\end{equation}
Finally, by defining a new dimensionless coordinate $x=\frac{r}{r_H}$, we obtain,
\begin{equation}
\Delta {\partial }_x\left(\Delta {\partial }_x\right)R+\left[\Delta \left(\frac{\Delta }{4x^2}-\frac{1}{2x^3}-\frac{m_\ell^2}{x^2}-\varpi^2\right)+{\left(\frac{E r_H}{\hbar c} -\frac{m_\ell{\Omega }_H}{x^2}\right)}^2\right]R=0,
\end{equation}
where we have introduced dimensionless parameters,
\begin{equation}
\varpi= {\omega }_0 r_H, \ \ \ \Omega_H={\omega }_H r_H. \label{redef}
\end{equation}

\section{Normal Form} \label{AppendixB}
The so called ``Normal Form" of an ordinary differential equation is the form when an ordinary differential equation is solved explicitly for the highest derivative \cite{NIST}. One may start with a general form of a linear second order ordinary differential equation as follows,
\begin{equation}
    \frac{d^2y}{dx^2}+p(x)\frac{dy}{dx}+q(x)y=0. \label{general ODE}
\end{equation}
We continue with applying a homotopic-family transformation, i.e., by making a special form of substitution for $y(x)$, which aims to remove the first order derivative term as follows \cite{2420},
\begin{align}
y&=Y(x)e^{-\frac{1}{2}\int{p(x)}dx},\\
\frac{dy}{dx}&=\frac{dY}{dx}e^{-\frac{1}{2}\int{p(x)}dx}-\frac{1}{2}Ype^{-\int{p(x)}dx},\\
\frac{d^2y}{dx^2}&=\frac{d^2Y}{dx^2}e^{-\frac{1}{2}\int{p(x)}dx}-\frac{1}{2}\frac{dY}{dx} pe^{-\frac{1}{2}\int{p(x)}dx}\nonumber \\
&~~~~-\frac{1}{2}Y\frac{dp}{dx}e^{-\frac{1}{2}\int{p(x)}dx}+\frac{1}{4}Yp^2e^{-\frac{1}{2}\int{p(x)}dx}.
\end{align}
After substituting the expressions to \eqref{general ODE}, we obtain second order differential equation without the first order derivative,
  \label{Normal Form}
\begin{gather}
    \frac{d^2Y}{dx^2}+\left(-\frac{1}{2}\frac{dp}{dx}-\frac{1}{4}p^2+q\right)Y=0. \label{normalform}
\end{gather}
The normal form is extremely useful to understand and estimate the physical nature of the ordinary differential equation. Suppose $Q(x)=-\frac{1}{2}\frac{dp}{dx}-\frac{1}{4}p^2+q>0$, the normal form reads $\frac{d^2Y}{dx^2}=-Q(x)Y$. Suppose we start with the case where $Y$ is positive, then, the second derivative will be negative and vice versa. It is clear that $Y$ will cross $x$-axis. If $Q(x)>0$ and $\int_1^\infty Q(x) dx=\infty$, then $Y(x)$ has infinitely many zeros on the positive $x$-axis. If $Q(x)<0$ then $Y(x)$ does not oscillate at all and has at most one zero  \cite{2420}.

\section{The Confluent Heun Equation and Its Solutions}\label{AppendixC}
The confluent Heun differential equation is a linear second order ordinary differential equation having the following canonical form \cite{Heun},  
\begin{align}
    \frac{d^2\psi_H}{dx^2}+\left(\alpha +\frac{\beta +1}{x}+\frac{\gamma +1}{x-1}\right)\frac{d\psi_H}{dx}+\left(\frac{\mu }{x}+\frac{\nu }{x-1}\right)\psi_H=0,
\end{align}
where,
\begin{align}
\mu&=\frac{1}{2}\left(\alpha-\beta-\gamma+\alpha\beta-\beta\gamma\right)-\eta,\\
\nu &=\frac{1}{2}\left(\alpha +\beta +\gamma +\alpha \gamma +\beta \gamma \right)+\delta +\eta.
\end{align}
The solutions are constructed in the terms of two independent confluent Heun functions as follow,
\begin{equation}
\psi_H=A\operatorname{HeunC}\left(\alpha ,\beta ,\gamma ,\delta ,\eta ,x\right)+Bx^{-\beta}\operatorname{HeunC}\left(\alpha ,-\beta ,\gamma ,\delta ,\eta ,x\right). \label{canonincalheun}
\end{equation}
There is a known polynomial condition for the confluent Heun function to become an $n_r-th$ order polynomial is given as follows,
 \begin{equation}
         \frac{\delta}{\alpha}+\frac{\beta +\gamma}{2}+1=-n_r,\quad n_r\in\mathbb{Z}. \label{HeunPol}
 \end{equation}
One can derive the following confluent Heun's differential equation in its normal form by recognizing $p$ and $q$ functions (following Appendix \ref{AppendixB}),
\begin{align}
p&=\alpha +\frac{\beta +1}{x}+\frac{\gamma +1}{x-1} \ \ , \ \  q=\frac{\mu }{x}+\frac{\nu }{x-1}, \\
\psi_H&=\Psi_H (x)e^{-\frac{1}{2}\alpha x}x^{-\frac{1}{2}\left(\beta+1\right)}{\left(x-1\right)}^{-\frac{1}{2}\left(\gamma +1\right)}.
\end{align}
Now, we evaluate the normal form's $ -\frac{1}{2}\frac{dp}{dx}-\frac{1}{4}p^2+q$ as follows,
\begin{align}
    -\frac{1}{2}\frac{dp}{dx} &= \frac{1}{x^2}\left(\frac{\beta +1}{2}\right)+\frac{1}{{\left(x-1\right)}^2}\left(\frac{\gamma +1}{2}\right), \\
-\frac{1}{4}p^2 &= -\frac{\alpha^2}{4}-\frac{1}{x^2}\left(\frac{\beta^2+1+2\beta}{4}\right)-\frac{1}{{\left(x-1\right)}^2}\left(\frac{\gamma^2+1+2\gamma}{4}\right) \nonumber \\
&~~~~-\frac{2}{x}\left(\frac{\alpha \beta +\alpha}{4}\right)-\frac{2}{x-1}\left(\frac{\alpha \gamma +\alpha}{4}\right)-\frac{2}{x\left(x-1\right)}\left(\frac{\beta \gamma +1+\beta +\gamma}{4}\right).
\end{align}
Combining all terms, we obtain the confluent Heun equation's normal form as follows,
\begin{equation}
    \frac{d^2\Psi_H}{dx^2}+\left(-\frac{\alpha^2}{4}+\frac{\frac{1}{2}-\eta}{x}+\frac{\frac{1}{4}-\frac{\beta^2}{4}}{x^2}+\frac{-\frac{1}{2}+\delta +\eta}{x-1}+\frac{\frac{1}{4}-\frac{\gamma^2}{4}}{{\left(x-1\right)}^2}\right)\Psi_H=0.
\end{equation}
The general solution to this equation can be expressed in term of confluent Heun function. This reads
\begin{multline}
    \Psi_H=e^{\frac{1}{2}\alpha x}x^{\frac{1}{2}}{\left(x-1\right)}^{\frac{1}{2}\left(\gamma +1\right)}\times\\\left[A_+ x^{+\frac{\beta}{2}}\operatorname{HeunC}\left(\alpha ,+\beta ,\gamma ,\delta ,\eta ,x\right)+A_- x^{-\frac{\beta}{2}}\operatorname{HeunC}\left(\alpha ,-\beta ,\gamma ,\delta ,\eta ,x\right)\right].
\end{multline}
However, it is convenient to write the solution in the following compact notation,
\begin{equation}
    \Psi_H=e^{\frac{1}{2}\alpha x}x^{\frac{1}{2}}{\left(x-1\right)}^{\frac{1}{2}\left(\gamma +1\right)}\left[A_{\pm}x^{\frac{\beta_\pm}{2}}\operatorname{HeunC}\left(\alpha ,\beta_\pm ,\gamma ,\delta ,\eta ,x\right)\right], \label{HeunSol}
\end{equation}
where the sign of $\beta_\pm=\pm\beta$ is to be determined by taking the desired boundary conditions into account.

\section{The Confluent Hypergeometric Equation}\label{AppendixD}
The confluent hypergeometric differential equation is a linear second order ordinary differential equation having this following form \cite{Bell},  
\begin{gather}
    \frac{d^2\psi_C}{dx^2}+\left(-\frac{1}{4}+\frac{k}{x}+\frac{\frac{1}{4}-m^2}{x^2}\right)\psi_C=0. \label{normalformhypergeometric}
\end{gather}
The solutions can be expressed in the terms of two independent Whittaker confluent hypergeometric functions, $M_{k,m}(x)$, as follow,
\begin{equation}
\psi_C=A M_{k,m}(x) +B M_{k,-m}(x).
\end{equation}
The Whittaker confluent hypergeometric functions are connected with the confluent hypergeometric functions, $F_{1,1}\left(a,b,x\right)$, as follows,
\begin{gather}
 M_{k,\pm m}=x^{\frac{1}{2}\pm m}e^{-\frac{x}{2}}F_{1,1}\left(\frac{1}{2}-k\pm m,1\pm 2m,x\right).
\end{gather}
In addition, the confluent hypergeometric function in an asymptotic region i.e., $x\to\infty$ can be written as
\begin{multline}
F_{1,1}\left(a,b,|x|\to\infty\right)=\frac{\Gamma(b)}{\Gamma(a)}e^xx^{a-b}F_{2,0}\left(b-a,1-a,-\frac{1}{x}\right)\\+  \frac{\Gamma(b)}{\Gamma(b-a)}(-x)^{-a}F_{2,0}\left(a,a-b+1,-\frac{1}{x}\right), \label{faratfar}
\end{multline}
where $F_{2,0}\left(a,b,x\right)$ is the hypergeometric function type $\{2,0\}$ \cite{NIST}.

\section{The Gauss Hypergeometric Equation}\label{AppendixE}
The Gauss hypergeometric differential equation is a linear second order ordinary differential equation having this following canonical form \cite{Bell},  
\begin{gather}
    x(1-x)\frac{d^2\psi_G}{dx^2}+\left(a_3-\left(a_1+a_2+1\right)x\right)\frac{d\psi_G}{dx}-a_1a_2\psi_G=0. 
\end{gather}
The solutions of the above equation are given in the terms of the Gauss hypergeometric functions, $F_{2,1}(a_1,a_2,a_3,x)$, as follow,
\begin{equation}
\psi_C=A F_{2,1}(a_1,a_2,a_3,x) +B x^{1-a_3} F_{2,1}(a_1-a_3+1,a_2-a_3+1,2-a_3,x).
\end{equation}
The normal form of the Gauss hypergeometric differential equation and its solutions are given as the following,
\begin{gather}
 \frac{d^2\Psi_G}{dx^2}-\left[\frac{x^2 \left\{(a_1-a_2)^2-1\right\}-2 x \left\{(a_1+a_2-1)a_3-2a_1 a_2\right\} a_3+a_3(a_3-2) }{4 (x-1)^2 x^2}\right]\Psi_G=0, \label{normalformgauss}\\
\Psi_G= x^{\frac{a_3}{2}}(1-x)^{\frac{1}{2}\left(1+a_1+a_2-a_3\right)} \psi_G.
\end{gather}
The following linear second order differential equation also has solutions in the terms of the Gauss hypergeometric functions,
\begin{gather}
    x^2(x+1)^2\frac{d^2\psi_G}{dx^2}+x(x+1)\frac{d\psi_G}{dx}+\left\{\mathcal{A}+\mathcal{B}x+\mathcal{C}x^2\right\}\psi_G=0.\label{gauss1}
\end{gather}
The general solution is 
\begin{multline}
 \psi_G= (x+1)^{1-\sqrt{1-\mathcal{A}+\mathcal{B}-\mathcal{C}}}\left[A x^{i\sqrt{\mathcal{A}}} F_{2,1}\left( \mathcal{A}_1,\mathcal{A}_2,\mathcal{A}_3,-x     \right)\right.\\ \left.+B x^{-i\sqrt{\mathcal{A}}} F_{2,1}\left( \mathcal{A}_4,\mathcal{A}_5,\mathcal{A}_6,-x     \right)\right],
\end{multline}
where
\begin{align}
   \mathcal{A}_1&=\frac{1}{2}\left(1+\sqrt{1-4\mathcal{C}}\right)+i\sqrt{\mathcal{A}}-\sqrt{1-\mathcal{A}+\mathcal{B}-\mathcal{C}},\\
   \mathcal{A}_2&=\frac{1}{2}\left(1-\sqrt{1-4\mathcal{C}}\right)+i\sqrt{\mathcal{A}}-\sqrt{1-\mathcal{A}+\mathcal{B}-\mathcal{C}},\\
   \mathcal{A}_3&=1+2i\sqrt{\mathcal{A}},\\
    \mathcal{A}_4&=\frac{1}{2}\left(1+\sqrt{1-4\mathcal{C}}\right)-i\sqrt{\mathcal{A}}-\sqrt{1-\mathcal{A}+\mathcal{B}-\mathcal{C}},\\
    \mathcal{A}_5&=\frac{1}{2}\left(1-\sqrt{1-4\mathcal{C}}\right)-i\sqrt{\mathcal{A}}-\sqrt{1-\mathcal{A}+\mathcal{B}-\mathcal{C}},\\
   \mathcal{A}_6&=1-2i\sqrt{\mathcal{A}}.
\end{align}
The Gauss hypergeometric function has a very useful property that allow us to probe the solution in an asymptotic limit,
\begin{align}
F_{2,1}(a_1,a_2,a_3,x) &= \frac{\Gamma(a_2-a_1)\Gamma(a_3)}{\Gamma(a_2)\Gamma(a_3-a_1)} (-x)^{-a_1}F_{2,1}\left(a_1,a_1-a_3+1,a_1-a_2+1,\frac{1}{x}\right) \nonumber  \\
&~~~+\frac{\Gamma(a_1-a_2)\Gamma(a_3)}{\Gamma(a_1)\Gamma(a_3-a_2)}(-x)^{-a_2} F_{2,1}\left(a_2,a_2-a_3+1,a_2-a_1+1,\frac{1}{x}\right) . \label{connection}
\end{align}
\end{appendices}

\bibliography{sn-bibliography}

\end{document}